\documentclass[10pt,a4paper]{article}
\usepackage[a4paper, left=2.0cm, right=2.0cm, top=2cm]{geometry}
\usepackage[normalem]{ulem}
\usepackage[utf8]{inputenc}
\usepackage{subfig}
\usepackage{amsmath, amsthm, amssymb, amsfonts}     
\usepackage{mathrsfs}                               
\usepackage{accents}                                
\usepackage{natbib}                                 
\usepackage[hidelinks]{hyperref}                    
\usepackage{doi}                                    
\usepackage{authblk}                                
\usepackage{siunitx}                                
\usepackage{booktabs}                               
\usepackage{multirow}                               
\usepackage{microtype}                              
\usepackage{makecell}                               
\usepackage{graphicx}
\usepackage{pgfplots}
\pgfplotsset{compat = newest}
\usepgflibrary{plotmarks} 
\usepackage{xcolor,soul}
\sethlcolor{lightgray}
\usepackage{orcidlink}

\let\savetabular\tabular
\def\tabular{\footnotesize\baselineskip=12pt\savetabular}

\title{Data-driven pressure field prediction for ships in regular sea states} 
\author[1]{Malte Loft \orcidlink{0000-0002-2820-3761}}
\author[1]{Henning Schwarz \orcidlink{0009-0000-0253-8902}}
\author[1]{Thomas Rung \orcidlink{0000-0002-3454-1804}}

\affil[1]{Institute for Fluid Dynamics and Ship Theory, Hamburg University of Technology, Am Schwarzenberg-Campus 4, D-21073 Hamburg, Germany}

\begin{document}

\maketitle


\begin{abstract}
Merchant shipping is responsible for more than 90\% of the global trade and has a significant environmental impact, accounting for over 2\% of global greenhouse gas emissions. Therefore, fuel-saving measures are becoming increasingly important in reducing the ecological footprint and increasing the fuel efficiency of maritime transport. Routing optimization systems, which require a rapid prediction of ambient-dependent fuel consumption,  represent an essential pillar here, e.g. to reduce added resistances due to seaways and/or wind. 
The paper aims to predict the added resistance due to seaways. In contrast to conventional methods the goal is achieved by surrogate modeling of the entire pressure fields.
To this end, an online/offline-procedure is applied to an exemplarily free-floating container  vessel. The online approach to be trained consists of two building blocks, namely a convolutional autoencoder (CAE)-based order reduction step and a neural network-based (NN) regression step that links the reduced space of the autoencoder with three control parameters that describe the sea state (wave height/steepness, encounter angle and wave length). 
Training data is obtained from time-averaged values for simulating instantaneous ship motion and pressure fields.
During the offline phase, the combined CAE/NN is trained to capture the time-averaged pressure fields for a variety of sea-state conditions. 
During ship operation (online phase), the surrogate model predicts the three-dimensional pressure fields in response to sea state conditions, projects the pressure fields onto the ship hull, and  integrates the corresponding resistances to guide the route.
The evaluation of the method shows promising results for the different building blocks and the concept could therefore represent an attractive approach for cost-effective surrogate modeling of complex multiphase flow fields.

\end{abstract}

\section{Introduction}
\label{sec:introduction}
Global warming is one of the central challenges in the 21st century, and massive reductions in greenhouse gas emissions are the most important strategy to counteract this challenge. Transportation, especially shipping, accounts for significant fractions of global CO$_2$, NO$_x$ and SO$_x$ emissions, and efforts to reduce fuel consumption are constantly increasing. In addition to various baseline optimizations, for example, with regard to the hydrodynamic performance of the shape of the hull (\cite{voith22}), route optimization systems are gaining importance. 
The additional resistance of ships caused by waves (\cite{liu2011prediction}) can be an important starting point for such strategies. 

Until now, route optimization systems lacked access to detailed hydrodynamic field data from the respective vessel. Thanks to the 
the now widely available hardware and software for machine learning (ML) procedures, digital field-data twins will be routinely established in the near future. 
Recently, increasing amounts of research have focused on (re)using simulation data to train ML-based surrogate models from CFD simulations, see for example \cite{eivazi:2020, wu:2021, pache2022data, hou2022novel, solera-rico:2024, schwarz:2024,Mi31122025}. 
Many of these strategies are based on neural networks and perform nonlinear dimension reductions.  
Although linear model reduction approaches such as proper orthogonal decomposition (POD; \cite{lumley1967structure}) are still popular due to the inherent interpretability and orthogonality of modes, autoencoders usually result in more compact latent spaces and more accurate reconstructions (\cite{milano2002neural, pache2022data}). Current research on the orthogonalization and interpretation of the latent space of AI-based reduction methods is showing initial success (\cite{solera-rico:2024, schwarz2025disentangled}). 

The paper proposes a surrogate model to predict the sea-state related drag from reconstructed pressure fields. 
A  convolutional autoencoder (CAE) is employed to reduce the dimension of the time-averaged, three-dimensional pressure fields obtained from CFD simulations.
The CAE is supplemented by a neural network (NN) which is used to map the seaway-related input parameters to the reduced representation. The combination enables a rapid online prediction of pressure fields and conceptually agrees with recent suggestions of  \cite{Swischuk.2019,agostini2020exploration, pache2022data, lazzara:2022}. 
The remainder of the paper is structured as follows: Section 
\ref{sec:hydrodynamics} outlines the hydrodynamic problem at hand. 
 Section \ref{sec:theory} discusses the mathematical models, that is, the CFD simulations and the data processing. The ML methods employed are described in Section \ref{sec:machinelearning}.
Results for a frequently studied container vessel are shown in Section \ref{sec:results} and conclusions are drawn in Section \ref{sec:conclusion}. The publication employs Einstein’s summation convention for lower-case
Latin subscripts. Vectors and tensors are defined with reference to Cartesian spatial coordinates $x_i$.

\section{Hydrodynamic Problem} 
\label{sec:hydrodynamics}

Merchant ships are exposed to a variety of fluid dynamic phenomena. The total resistance during ship operation is influenced, among other things, by parasitic factors due to waves, wind, currents, and fouling of the ship's coating.
Previous work focused on reproducing wind-induced forces on the superstructure of a container vessel using ML-based and POD-based surrogate models, cf. \cite{pache2022data}. In contrast, this study  investigates the influence of waves on the resistance of a free-floating vessel. 

\subsection{Investigated Vessel} 
\label{sec:KRISO}
\begin{figure}[bh!]
    \centering
    \includegraphics[width=0.75\textwidth]{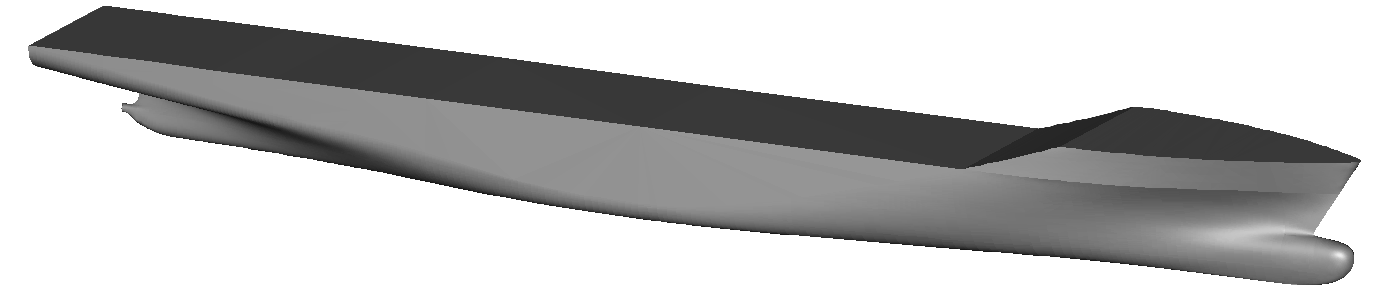}
    \caption{Investigated KRISO container vessel ($\Lambda=31.599$, $L_\mathrm{pp}=7.279\,\si{m}$, $\text{Re} = 1.6\cdot 10^7$, $\text{Fn} = 0.26$).}
    \label{fig:KRISO}  
\end{figure}

The geometry used in this study is based on the KRISO container vessel (KCS, cf. \cite{KCS}) at model scale $\Lambda=31.599$, cf. Figure \ref{fig:KRISO}. The cruising speed is set to $u_{1} = 2.1962\,\si{m\per s}$, which yields a Reynolds number of $\text{Re}= 1.6\cdot 10^7$. It is assumed that the additional forces are pressure dominated and wave-induced changes of friction forces can be neglected. This assumption is confirmed by the simulation data shown in Table \ref{tab:frictionpressure}. Moreover, the motion of the vessel is restricted to heave, pitch, sway and surge during the current simulations.

\subsection{Investigated Scenario} 
\label{sec:procedure}
\begin{figure}[ht!]
    \centering
    \hspace*{\fill}%
    \subfloat[Bird view.]{\includegraphics[scale=1]{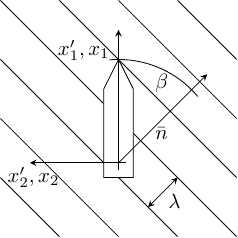}}\hfill%
    \subfloat[Side view.]{\includegraphics[scale=1]{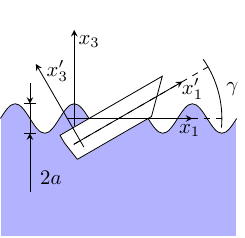}}%
    \hspace*{\fill}%
    \caption{Illustration of the investigated scenarios and coordinate systems characterized by the global coordinates ($x_i$), the local ship-fixed coordinates ($x'_i$), the angle $\beta$ between the direction $\bar{n}$ of the  propagating  waves and the global $x_1$ coordinate as well as the pitch angle $\gamma$.}
    \label{fig:coordinates}  
\end{figure}

The surrogate model aims to reconstruct the time-averaged, three-dimensional deviation of the 
pressure field from the corresponding calm-water pressure field in response to sea-state conditions. 
In the current study, the sea state is represented by three parameters: (I) the wave length $\lambda$, (II) the wave height $2a$ and (III) the incidence/encounter angle of the waves $\beta$. Figure \ref{fig:coordinates} illustrates the three control parameters and outlines the two coordinate systems employed during this study, i.e., a ship following system (${x}'_i$) and the
inertial system ($x_i$). 
The angle between the global ($x_1$-) and local ($x'_1$-) longitudinal axis refers to the pitch angle $\gamma$. 
The incidence angle $\beta$ of encountering waves is defined as the angle between the propagation direction of the waves and the global $x_1$-axis. 
The focus of this study is the drag coefficient in the inertia coordinates ($C_\mathrm{D}$) and the  ship-fixed coordinates ($C_\mathrm{D}'$). These normalize the force $F_1$ or $F_1'$, with the density of the water $\rho$, the square of the cruising speed $u_1$, and a reference surface $S$, viz.
\begin{equation}
   C_\mathrm{D} = \frac{F_1}{\rho u_1^2 S} \qquad \text{and} \qquad C'_\mathrm{D} = \frac{F'_1}{\rho u_1^2 S}. 
   \label{eq:drag}
\end{equation}
Emphasis is given to the corresponding additional drag, 
which is defined as the difference between the drag experienced in a wave scenario and the calm water resistance, e.g.,  $\Delta C_\mathrm{D} = C_\mathrm{D} - C_\mathrm{D,0}$.

\smallskip
The study considers $150$ discrete seaway conditions, including five normalized wave lengths $\lambda/L_{pp}$, three normalized wave heights $a/L_{pp}$ and ten different wave encounter angles $\beta$. Table \ref{tab:parameterspace} provides an overview of the parameter space considered. The parameters used are chosen to ensure a good compromise between a moderate computational effort and an adequate coverage of a broad and realistic parameter space. 
A brief notation is defined in which a letter denotes the angle $\beta$ (a-j) followed by two digits, i.e. one for the wave length (1-5) and one for the height of the waves (1-3), e.g., b4-2: $[\beta = 20\,\text{deg, }\lambda = 1.25\, L_\mathrm{pp}, 2a = 1/100 \,L_\mathrm{pp}]$. 

\begin{table}[ht!]
    \caption{Overview of the discrete parameter spectrum.}
    \centering
    \begin{tabular}{cccc}
        \toprule
        Parameter & Notation & Range & Number of parameters $n_i$ \\ \midrule 
        Incidence angle $\beta$  &$[a,...,j]$ &  $[0, 20, 40, 60, 80, 100, 120, 140, 160, 180 ]\, \text{deg}$ & $10$ \\
        Wave length $\lambda$  &$[1,...,5]$& $[0.50, 0.75, 1.00, 1.25, 1.50]\,L_\mathrm{pp}$ & $5$ \\ 
        Wave height $2a$ & $[1,2,3]$& $[1/200, 1/100, 1/66 ]\,L_\mathrm{pp}$ & $3$ \\
        
        \midrule
        Sum $\sum$  & & $n_\lambda \cdot n_{2a} \cdot n_\beta$ & $150$\\
        \bottomrule
    \end{tabular}
    \label{tab:parameterspace}
\end{table}

 As the table shows, unsteady CFD simulations were performed for angles of incidence of $0\deg\le \beta \le 180\deg$. Each of the ten angles of incidence considered contains 15 different wave fields. To also account for angles of incidence in the range $200\deg\le\beta\le360\deg$, the pressure fields for the nine angles $0\deg\le\beta\le160\deg$ are mirrored,
 i.e., pressure fields for $\beta = 20\deg$ are flipped along the $x'_2$-axis to obtain pressure fields for $\beta = 340\deg$, etc. The dataset used by the ML procedure therefore contains a total of 285 cases.

\subsection{Machine Learning Approach} 
\label{sec:onoff}
During the offline phase, the combined CAE/NN is trained to learn the pressure data obtained for the cases outlined in Table \ref{tab:parameterspace}. The entire process contains four building blocks: (a) projection of CFD-based pressure fields, obtained on unstructured locally refined meshes, into a structured grid environment, (b) time-averaging of pressure fields, (c) machine learning of pressure fields, and (d) transferring surrogate model predictions to the vessel geometry to evaluate the (added) resistance, cf. Eq.~\eqref{eq:drag}. 

\begin{figure}[ht!]
    \centering
    \subfloat[Time averaged data as input for machine learning.]{\includegraphics[width=0.50\textwidth]{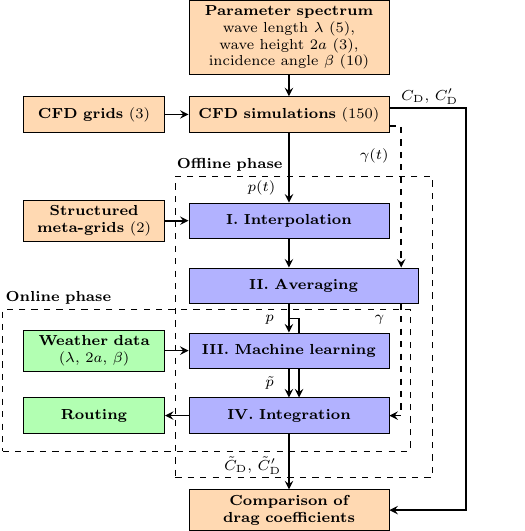} \label{fig:process1}} 
    \subfloat[Instantaneous data as input for machine learning.]{\includegraphics[width=0.50\textwidth]{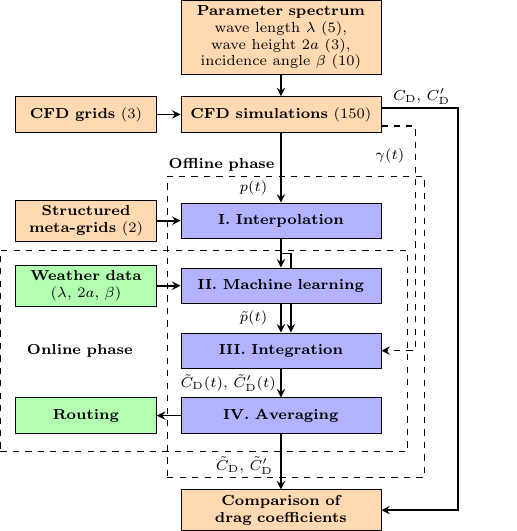} \label{fig:process2}}
    \caption{CFD simulations are performed due to the according parameter spectrum (top). Three different unstructured CFD grids are generated for the simulations. The surrogate models always consists of four major steps: interpolation, averaging, machine learning and integration. Two different surrogate modeling procedures are represented: (a) Time averaged data are used for the machine learning step, Integration is done as a last step in consideration of an averaged pitch angle. (b) Machine learning is done with instantaneous pressure fields, integration is done for instantaneous data and averaging is moved to the last step of the procedure. The dashed line of the pitch angle $\gamma$ indicates, that this information is only used for the integration step in global coordinates since the interpolated pressure fields are written in the local ship coordinate system. 
    }
    \label{fig:process}  
\end{figure}

Due to the unsteady nature of the CFD simulations, two different approaches are conceivable, cf. Figures \ref{fig:process1} and \ref{fig:process2}. 
Both procedures differ only in the order of these buildings blocks, in particular whether time-averaged or time-dependent fields are learned. In line with previous studies, a more convenient strategy is to train the surrogate model based on time-averaged fields. For this approach, step (d) is performed last and the averaging is performed prior to the learning process.  Since only spatial dependencies and features are considered, the latent space is expected to be low dimensional. 
A challenge of this strategy concerns the inherent ship motion and the disadvantage of this approach becomes apparent when the pressures extracted from the surrogate model in ship-fixed coordinates are to reconstruct the forces in global coordinates, which would require a matching pitch angle.
In this case, the second approach seems more reasonable, cf. results displayed in Section \ref{sec:verifyDataProcessing}. Here, the surrogate model is trained based on the instantaneous pressure fields. However, the second strategy would increase the training data volume by several orders of magnitude and is therefore discarded in the present study.

\section{Computational Fluid Dynamics Model}
\label{sec:theory}

The CFD simulations use an implicit second-order finite volume method as described by \cite{rung2009challenges} and \cite{volkner2017analysis}. The algorithm is based on the strong conservation form of the momentum equations. A pressure correction algorithm is implemented to determine the pressure for incompressible flows. For the spatial discretization, unstructured grids based on arbitrary polyhedral cells are used together with a cell-centered, collocated arrangement of the variables. 
Temporal derivatives are approximated by an implicit three time level method and spatial integrals by the midpoint rule. Second order hybrid central/upwind approximations are used to approximate the convective fluxes. In addition to the momentum equations, a well known Volume of Fluid (VoF) approach is used in combination with a high order interface scheme (HRIC) to model the dynamic air-sea interface (\cite{muzaferija1999two}).
Turbulence is closed by a basic RANS $k$-$\omega$-MSST model (\cite{menter1994two}). 
The original six Degree of Freedom (DoF) system is reduced to a 4DoF system, that allows a free heave, pitch, sway and surge motion of the ship. Yaw and roll motions are suppressed, because they lead to undesired resonance behavior and an highly increased effort to automatically schedule the total number of simulations.  
Equations of motion are solved with a quaternion-based motion approach, as previously implemented by \cite{luo2017computation}.
The ship stays on course with the help of a virtual spring system, that is adjusted to the individual operational parameters of each simulation run and coupled with the sway and surge motion of the ship. Spring stiffness is determined in such a way that the natural frequency of the spring mass system stays always significantly below the driven frequency of the wave. This is archived for the majority of all parameter combinations, with the exception of a few cases where the wave propagates with the same velocity relative to the ship. 
\begin{figure}[ht!]
    \centering
    \subfloat[$x'_1$-$x'_3$ section.]{\includegraphics*[width=0.49\textwidth, trim=0 0 0 0, clip=True]{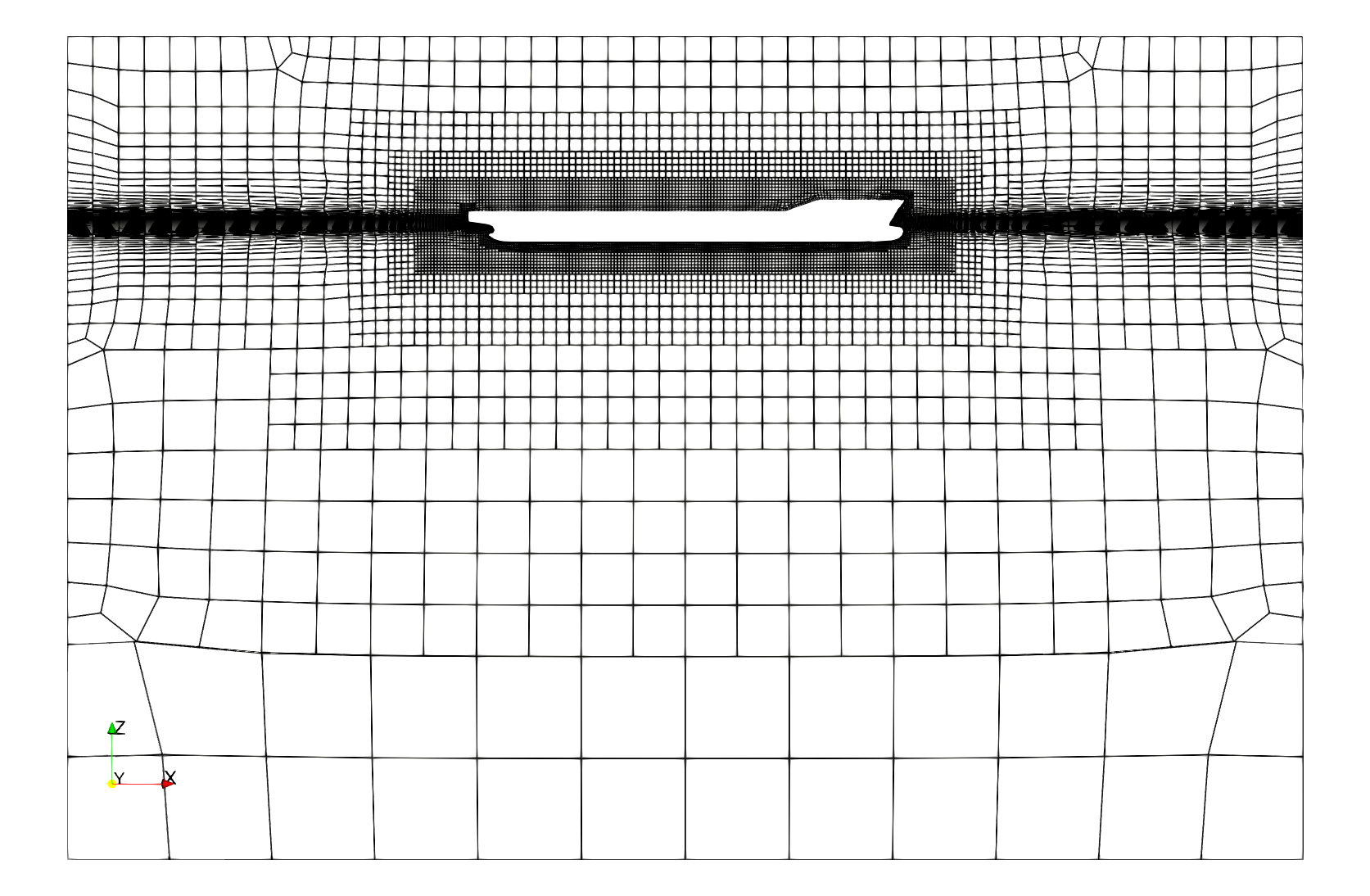}}
    \subfloat[$x'_2$-$x'_3$ section.]{\includegraphics*[width=0.47\textwidth, trim=0 0 0 0, clip=True]{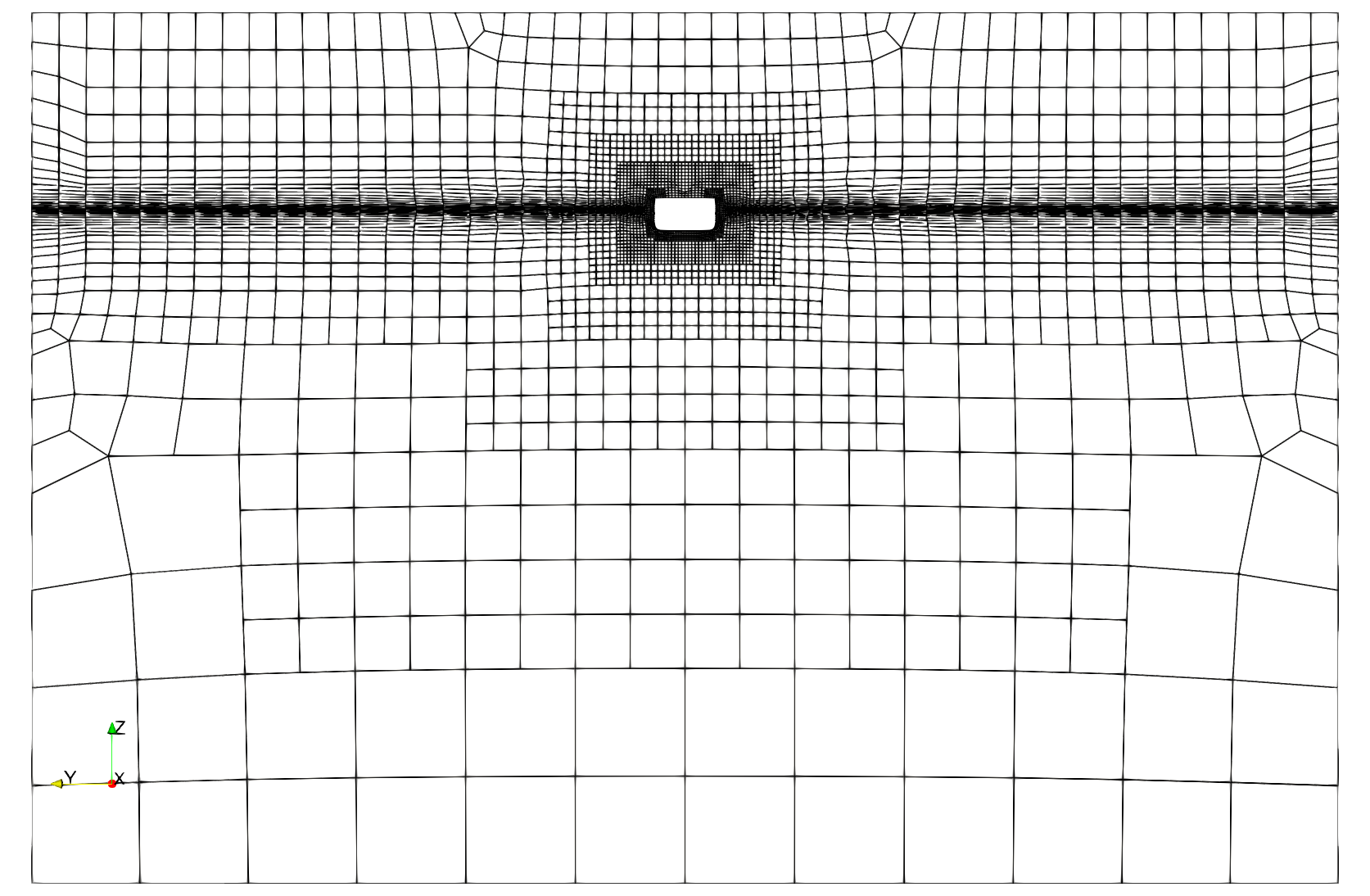}}
    \caption{Views of the unstructured CFD grid for the exemplary simulated KRISO container ship (KCS).}
    \label{fig:meshes}
\end{figure}
For all $150$ parameter combinations, a total of 3 different, unstructured grids are generated with $1.2$ to $3.1$ million cells. The domain has an extension of $3\,L_\text{pp} \times 3\,L_\text{pp} \times 2\,L_\text{pp}$ in local coordinate directions $x'_1$, $x'_2$ and $x'_3$. In line with \cite{luo2019numerical} arbitrary sea states are generated with an implicit forcing approach, that is based on linear wave theory. 

The resolution of the employed pitching CFD grids depends on the investigated wave length.
 This ensures that at least $16$ control volumes are available to resolve  the wave length. In the vertical direction, the resolution per wave height varies between $6$ and $18$ cells. 
Figure \ref{fig:meshes} depicts two different sections of an exemplary grid. Mind that the numerical grid is associated with the ship-fixed coordinate system and therefore the free surface can also pitch against the longitudinal axis.
To ensure a sufficient resolution of the free surface, the grid is refined in the $x'_1-x'_3$ plane. 
Due to the suppression of the roll motion, a refinement in the $x'_2-x'_3$ plane is not necessary. 
 The employed time-step size $\Delta t$ varies with the investigated wave frequency $f$, i.e., $f \, \Delta t = 10^{-3}$.  
Each simulation comprises a total of 20.000 time steps, where the last 10.000 time steps are reserved for data processing, with only every twentieth time step being taken into account. 
The domain is divided into approximately $250$ partitions and parallel simulations are performed using the message passing interface (MPI) communication protocol.

\subsection{Data Transfer}
\label{sec:dataprocessing}
The CFD simulations were performed on commonly used unstructured, locally refined grids. In contrast, the ML method uses structured grid data. This requires the definition of a data transfer procedure.

\begin{figure}[ht!]
    \centering
    \subfloat[Meta-grid A.]{\includegraphics*[width=0.49\textwidth, trim=0 0 0 0, clip=True]{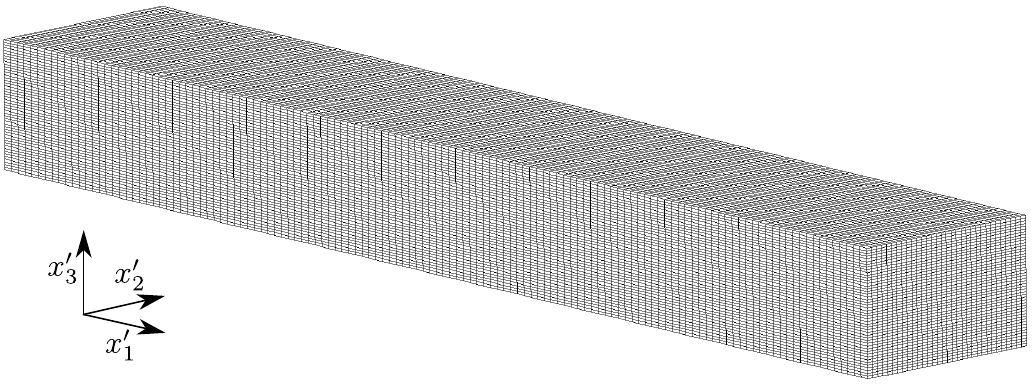}}
    \subfloat[Meta-grid B.]{\includegraphics*[width=0.47\textwidth, trim=0 0 0 0, clip=True]{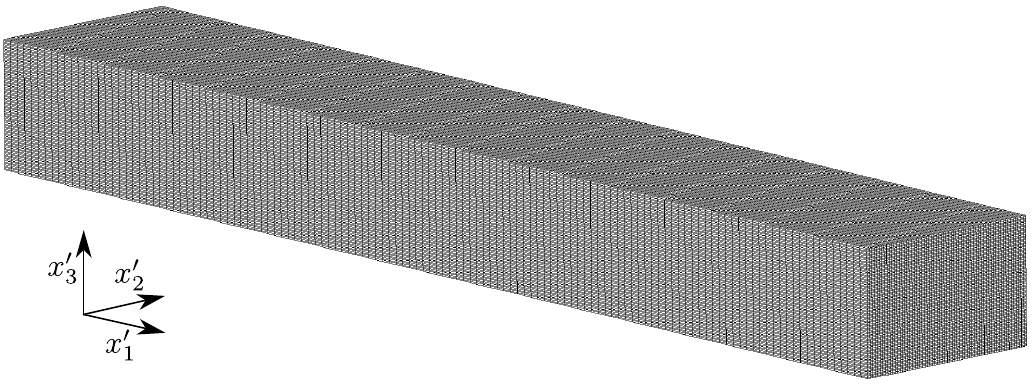}}
    \caption{Views of the employed two meta-grids.}
    \label{fig:metagrids}
\end{figure}

\begin{figure}[bh!]
    \centering
    \includegraphics[width=0.99\textwidth]{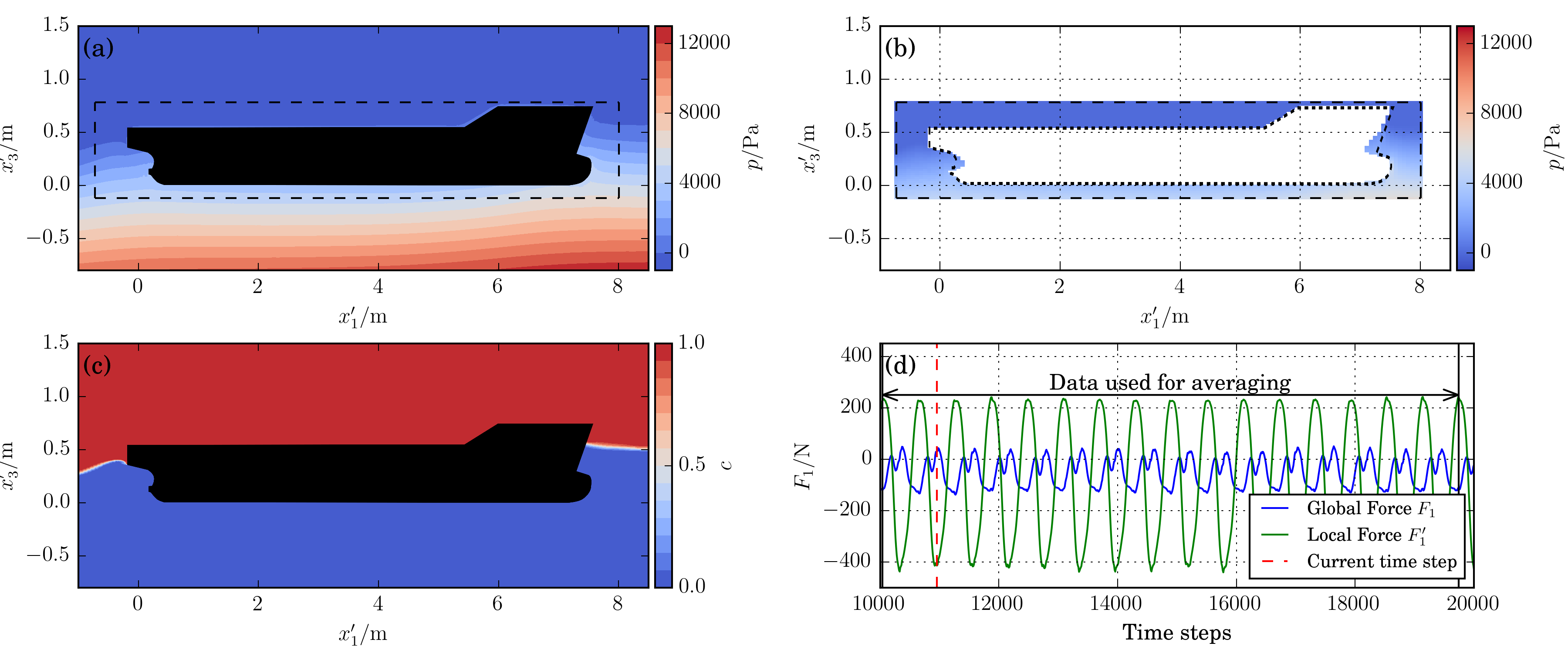}
    \caption{Comparison of the simulated (a) and interpolated (b, meta-grid A) instantaneous pressure fields of an exemplary simulation run (j3-3). 
    The corresponding instantaneous air concentration field obtained from the CFD simulation is shown in (c). The temporal evolutions of the CFD-predicted global $F_1$ and local $F'_1$ forces are displayed in (d).
    }
    \label{fig:simulation}
\end{figure}

The pressure fields computed by the CFD method on unstructured grids sketched in Figure \ref{fig:meshes} are projected into structured, orthogonal meta-grids covering the ship's interior and surrounding. These grids form the basis of the machine learning method. In the current study, two meta-grids (A, B) are employed, which consist of $128 \times 32 \times 48$ $(x'_1, x'_2 , x'_3)$ and $128 \times 64 \times 96$ $(x'_1, x'_2 , x'_3)$ cells, respectively. Both grids cover the same domain and only differ in the resolution of $\Delta x'_2$ and $\Delta x'_3$, cf. Figure \ref{fig:metagrids}.
The interpolation into the meta-grid is performed via an inverse distance approach, where the pressure values of the $5$ closest CFD cells are used to determine the pressure in each cell of the  meta-grid, cf. Figures \ref{fig:simulation} (a) and (b). Mind that locations inside the ship hull are marked and processed separately in the surrogate model, cf. Section \ref{sec:datasets}. 
Since the simulated number of effective wave periods strongly depends on the incidence angle $\beta$, a Hilbert transformation (\cite{loft2023twophase}) is performed to determine the time window of the maximum whole number of wave periods in the present data set, cf. Figure \ref{fig:simulation} (d). All data within this time window are used for the averaging respectively.

To integrate the drag forces from pressures predicted by the ML-based surrogate model on the meta-grid, the CFD discretization of the hull is re-considered and the nearest meta-grid node is identified for each surface element of the CFD grid.
Note that the hydrostatic pressure fields are included for the integration step, but the surrogate model only receives pressure differences.

\section{Machine Learning}
\label{sec:machinelearning}
The applied ML strategy consists of two parts: a  reduction of the meta-grid pressure fields using a convolutional autoencoder (CAE) and a regression of the three sea-state parameters into the reduced representation, in line with the approaches used in \cite{agostini2020exploration, Swischuk.2019, pache2022data, lazzara:2022}. The regression is performed with a fully connected neural network in this work, but other approaches such as a Gaussian process regression or polynomial regression are possible. 
All neural networks in this work are implemented using TensorFlow (\cite{TF:2016}) and Keras (\cite{chollet2015keras}).

\subsection{Data Sets}
\label{sec:datasets}
The $285$ cases are divided into three different data sets. The training set involves $249$ cases 
used to train the models. The validation set is used to terminate training early to avoid overfitting to the training data.  
 A final test dataset is used after training to test the full surrogate model shown in Figure \ref{fig:process}. Both the validation and test data sets contain data from 18 cases each.

As described in Section \ref{sec:dataprocessing}, two regular equidistant grids A and B are assessed  in this study. 
The meta-grid locations inside the ship can be handled in different ways. 
Two different strategies are tested in this regard. The first strategy simply assigns $zero$-values to these cells (Int.-Zero).
The second strategy iteratively propagates local information from the boundary of the ship into the interior: For a cell on the boundary of the ship, i.e., a cell with neighboring fluid cells, the mean of the pressures of these neighboring fluid cells is calculated. This process continues until all interior locations are filled with values (Int.-Mean).

The training, validation and test data is finally normalized to the interval $[0,1]$ using $\min$-$\max$ normalization. Pressure predictions by the surrogate model are easily mapped back to the original order of magnitude, with the data for interior points again deleted.

\subsection{Machine Learning Models}
An autoencoder comprises two neural networks, namely an encoder $\mathbf{E}$ and a decoder $\mathbf{D}$. The encoder reduces the dimension of the input by mapping it to a latent space of lower dimension. Using the low dimensional data as input, the decoder reconstructs the input in the full dimension. By training both components simultaneously, an autoencoder learns important low dimensional features that can be used to reconstruct the input. A convolutional autoencoder (CAE) uses convolutional layers in the encoder and the decoder. Due to the typically small size of the convolution kernel, this is helpful when considering physical data, as spatial dependencies can be captured. Moreover, convolutional layers typically require much fewer parameters than fully connected layers, which reduces storage capacity and helps to prevent overfitting. 
The structure of the CAE in this work is given in Table \ref{tab:cae}. We employ \textsf{LeakyReLU} (\cite{maas:2013}) activation functions with a slope parameter $\alpha=0.01$ to connect the layers. The last layers of the encoder and the decoder are linear.

\begin{table}[ht!]
\caption{CAE structure. Output shapes refer to meta-grids A[B]. All Conv3D and Conv3DTranspose layers use filters of size $3\times3$ and strides of 2.}
\centering
\begin{tabular}{cccc}
    \toprule
    Encoder Layer & Output shape & Decoder Layer & Output shape\\
    \midrule
 Input & (32[64],128,48[96],1) & Dense(3072[12288]) & (3072[12288])\\
     Conv3D(filters=8) & (16[32],64,24[48],8) &  Reshape(2[4],8,3[6],64) & (2[4],8,3[6],64)\\
     Conv3D(filters=16) & (8[16],32,12[24],16) & Conv3DTranspose(filters=32) & (4[8],16,6[12],32) \\
     Conv3D(filters=32) & (4[8],16,6[12],32) & Conv3DTranspose(filters=16) & (8[16],32,12[24],16) \\
     Conv3D(filters=64) & (2[4],8,3[6],64) & Conv3DTranspose(filters=8) & (16[32],64,24[48],8) \\
     Flatten() & (3072[12288])  & Conv3DTranspose(filters=1) & (32[64],128,48[96],1) \\  
     Dense(10) & (10) &  \\
    \bottomrule
\end{tabular}
\label{tab:cae}
\end{table}
 
In line with the work of \cite{pache2022data}, the present study also employs a fully connected neural network to perform the parameter regression. The structure of the regression network is depicted in Table \ref{tab:regression_net}.
\begin{table}[ht!]
\caption{Regression network structure.}
\centering
\begin{tabular}{cc}
    \toprule
    Layer & Output shape \\
    \midrule
 Input & (3)\\
     Dense(200) & (200)\\
     Dense(200) & (200)  \\
     Dense(10) & (10)  \\
    \bottomrule
\end{tabular}
\label{tab:regression_net}
\end{table}
As for the CAE, \textsf{LeakyReLU} activation functions are used in combination with $\alpha=0.01$  for the hidden layers. The output layer is linear.

\subsection{Model Training}
\begin{figure}[ht!]
    \centering
    \includegraphics[scale=1]{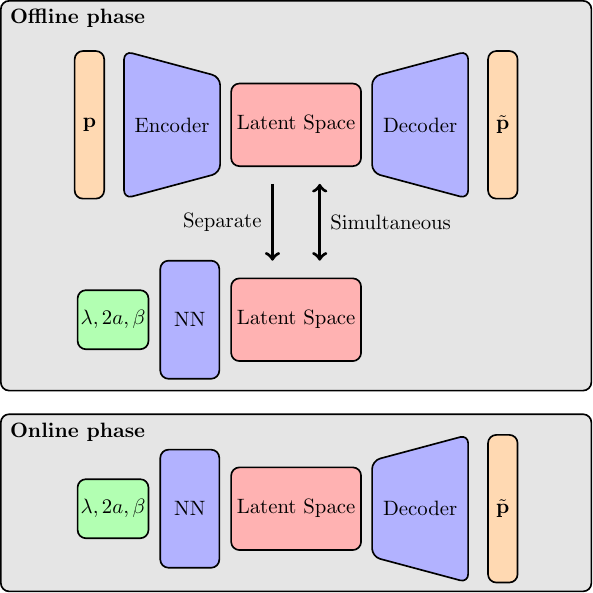}
    \caption{Offline and online phases of the neural network based surrogate model.}
    \label{fig:surrogate_model}
\end{figure}

For training the CAE and regression network, two options are conceivable: separate or simultaneous training of the two components. In the separate case, the CAE is first trained using the 3D pressure fields, and the resulting latent space is subsequently used to train the regression network. 
In the simultaneous case, the latent space is also influenced by the training of the regression network. Both approaches were tested in this work. 
The complete machine learning approach, including offline (training) and online (application) phases, is illustrated in Figure \ref{fig:surrogate_model}.
For the separate training, the CAE is trained with the Adam optimizer (\cite{Kingma:2015}) and a learning rate of $10^{-3}$. The mean squared error (MSE) is used for the loss function. 
\begin{align}
        \mathrm{MSE}(\mathbf{p},\tilde{\mathbf{p}})=\text{MSE}(\mathbf{p}, \mathbf{D}(\mathbf{E}(\mathbf{p}))) := \frac{1}{n}\lVert \mathbf{p}-\tilde{\mathbf{p}} \rVert^2_2, \quad \mathbf{p} \in \mathbb{R}^n, \tilde{\mathbf{p}} \in \mathbb{R}^n.
        \label{eq:mse}
\end{align}
Similar to the loss function used by \cite{pache2022data}, the MSE is set (manually) to zero for cells inside the ship, since there are no pressure values in these cells that can be predicted.
The mini batch size refers to 2. Larger mini batch sizes lead to poorer generalization to unseen data in this study. The training is stopped when the error on the validation set has not decreased for 50 epochs.
The regression network is subsequently trained using the MSE and Adam with a mini batch size of 128. Again, the training is stopped when the error on the validation set has not decreased for 50 epochs.

The loss function for the simultaneous training can potentially contain two more terms, i.e., an additional latent space loss for the regression network to enable a mapping from the input parameters to the reduced representation, supplemented by a prediction loss acting on the pressure field prediction obtained from the three sea-state parameters. The latter thus acts on the output of the layers that are applied in the online phase:
\begin{align}
    \label{eq:simultan_loss}
    \begin{split}
    \text{Loss}_\text{simultaneous}(\mathbf{p}, \lambda,2a,\beta, \tilde{\mathbf{p}}) &= \nu_{\text{reconst}} \cdot \text{MSE}(\mathbf{p}, \tilde{\mathbf{p}})\\ &+ \nu_\text{latent} \cdot \text{MSE}(\mathbf{E}(\mathbf{p}), \textbf{NN}(\lambda,2a,\beta))\\ &+ \nu_\text{predict} \cdot \text{MSE}(\mathbf{p},\textbf{D}\textbf{(NN}(\lambda,2a,\beta)))
    \end{split}
\end{align}

Again, the MSE contributions for the full-dimensional data are manually set to zero for cells inside the ship, the mini batch size refers to 2 and the Adam optimizer with a learning rate of $10^{-3}$ is employed. The training is also stopped when the error on the validation set has not decreased for $50$ epochs. 

The factors for the reconstruction and prediction losses in Eq. \eqref{eq:simultan_loss} are set to unit values with equal weight, i.e., $\nu_{\text{reconst}}=\nu_{\text{predict}}=1$.
 We have experienced that too large values for $\nu_\text{latent}$, e.g., $\nu_\text{latent}\ge 0.1$, 
 can cause problems in the training process, in which the latent space loss dominates. On the contrary, using only the reconstruction and prediction losses already provides satisfactory results. For that reason, $\nu_{\text{latent}}=0$ is set in this work. Without going into further detail, it is stated that setting $\nu_{\text{latent}}=10^{-3}$ gives similar results as $\nu_{\text{latent}}=0$.
The present study is confined to a simultaneous training method, which usually provides improved results (\cite{lusch:2018,wu:2021,schwarz:2024}).

\section{Results}
\label{sec:results}
The discussion of the results is limited to the analysis of the wave-induced resistance  of the considered ship. To this end, additional friction forces can be neglected, as the data in Table \ref{tab:frictionpressure} shows. Relative to the calm water resistance, the additional drag coefficient for all 150 applications examined is, on average, more than twenty times higher than the corresponding friction coefficient. This also applies to the maximum drag increase.   
\begin{table}[ht!]
\caption{Relative change of drag and friction coefficients ($\Delta C_\mathrm{D}$, $\Delta C_\mathrm{f}$) in comparison to the calm-water simulation ($C_\mathrm{D,0}$, $C_\mathrm{f,0}$) for the global coordinate system. Indication of the mean and maximum ratios for all $150$ CFD simulation runs.}
\centering
\begin{tabular}{cccc}
    \toprule
    Mean $|\Delta C_\mathrm{D}/C_\mathrm{D,0}|$ &  Max $|\Delta C_\mathrm{D}/C_\mathrm{D,0}|$ & Mean $|\Delta C_\mathrm{f}/C_\mathrm{f,0}|$ &  Max $|\Delta C_\mathrm{f}/C_\mathrm{f,0}|$ \\  \midrule 
    $16.8\,\%$ & $123.5\,\%$ & $0.7\,\%$ & $3.9\,\%$ \\
    \bottomrule
\end{tabular}
\label{tab:frictionpressure}
\end{table}

Each component of the surrogate model represents a potential source of error that reduces the accuracy of the entire process chain. To quantify these errors separately,
the results section is divided into three parts. First, data transfer influences on the reproducibility of the additional drag contributions are evaluated in Section \ref{sec:verifyDataProcessing}. For this purpose, the CFD pressure fields are transferred to the meta-grids as shown in Section \ref{sec:dataprocessing}. This reconstructs the surface pressures and integrates the drag forces without machine learning.
Second, the prediction quality of the full surrogate model is assessed by calculating the RMS errors of the predicted pressure fields and resistance.   Aspects investigated in Section \ref{sec:verifySurrogateModel} are the influences of (a) the values used inside the ship (Int.-Zero vs. Int.-Mean) and (b) the meta-grid. Finally, the influence of the training approach as well as the dimension of the latent space on the quality of the results is briefly discussed in Section \ref{sec:hyperTraining}.

\subsection{Verification of Data Transfer Influences}
\label{sec:verifyDataProcessing}
\begin{figure}[ht!]
    \centering
    \includegraphics[width=0.99\textwidth]{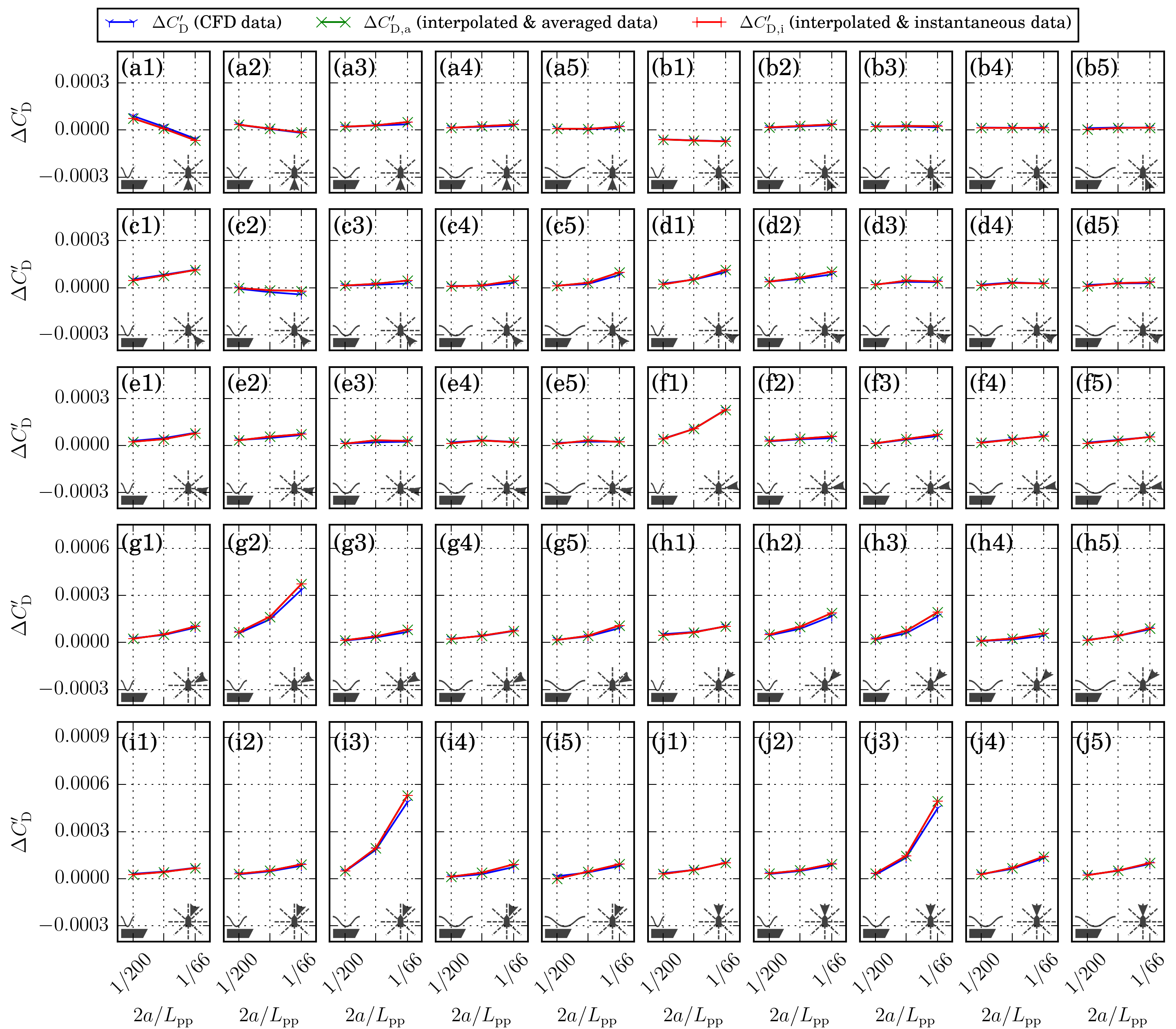}
    \caption{Analysis of the influences of the meta-grid data transfer by comparing the additional drag coefficients over the wave steepness ($2a/L_{pp}$) in \underline{ship-fixed coordinates} obtained from CFD simulations ($\Delta C'_\mathrm{D}$, blue) with results extracted from the transferred pressure fields generated by the time-averaged ($\Delta C'_\mathrm{D,a}$, green) and the instantaneous ($\Delta C'_\mathrm{D,i}$, red) strategy using meta-grid B. Bottom icons in each sub-figure indicate the respective wave-length (ascending from 1-5) and the respective wave encounter angle (ascending from a-j), see  Section \ref{sec:procedure}. 
    }
    \label{fig:results_interpolated_vs_cfd_local}  
\end{figure}

\begin{figure}[ht!]
    \centering
    \includegraphics[width=0.99\textwidth]{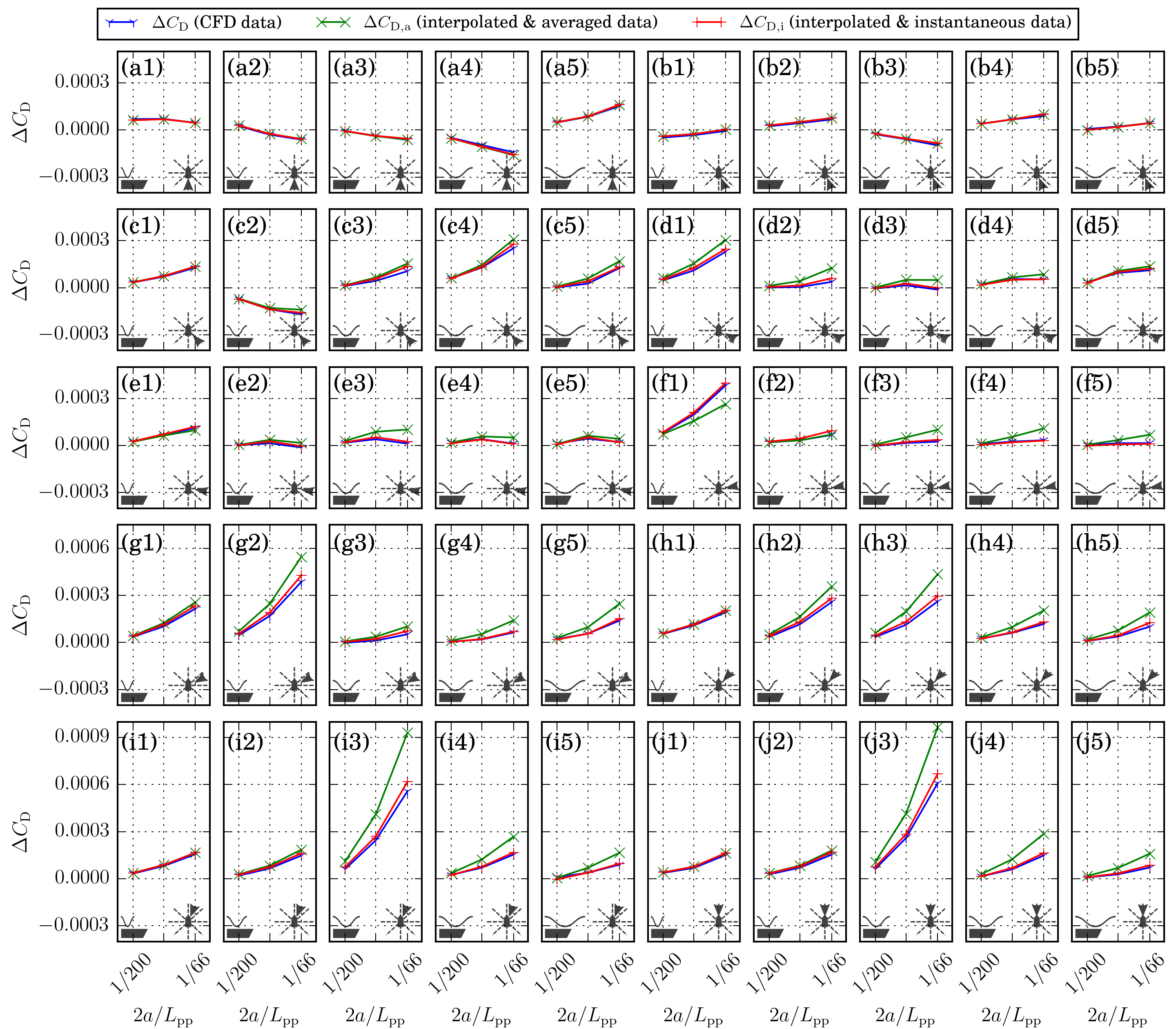}
    \caption{
    Analysis of the influences of the meta-grid data transfer by comparing the additional drag coefficients over the wave steepness ($2a/L_{pp}$) in \underline{inertia coordinates} obtained from CFD simulations ($\Delta C_\mathrm{D}$, blue) with results extracted from the transferred pressure fields generated by the time-averaged ($\Delta C_\mathrm{D,a}$, green) and the instantaneous ($\Delta C_\mathrm{D,i}$, red) strategy using meta-grid B. Bottom icons in each sub-figure indicate the respective wave-length (ascending from 1-5) and the respective wave encounter angle (ascending from a-j), see  Section \ref{sec:procedure}. 
    }
    \label{fig:results_interpolated_vs_cfd_global}  
\end{figure}

Figure \ref{fig:results_interpolated_vs_cfd_local} displays 
the drag coefficients in ship-fixed coordinates for the meta grid with the higher resolution, i.e., meta grid B. Similarly, Figure \ref{fig:results_interpolated_vs_cfd_global} shows the drag coefficients in inertia coordinates for the meta-grid B.
Both figures compare the results obtained from the time-averaged approach outlined in Figure \ref{fig:process1} (green) and the instantaneous approach outlined in Figure \ref{fig:process2} (red) with the original CFD data (blue).  Ordinate values depict the additional drag and abscissa values correspond to the wave height. All sub-figures provide icons that indicate the wave encounter direction \textendash{} from following seas (top left) to head seas (bottom right) \textendash{} and the wave length (ascending from 1-5).  The notation is explained in Section \ref{sec:procedure}. 

In the ship-fixed pitching coordinate system, the time-averaged and the instantaneous strategies are mathematically identical and their results agree. 
In general, drag coefficients are reproduced very well for all investigated scenarios. 
Small deviations due to the transfer of data into (and out of) the meta-grid are observed for the resonance head-sea cases (i3-3 or j3-3), where the wave length matches the ship length $L_\mathrm{pp}$ and the steepness is largest. Here localized high pressure amplitudes predicted by the CFD simulation are smoothed out during the transfer into the (locally) coarser meta-grids, and therefore, pressure fields deviate from the CFD simulation. Nevertheless, the average resistance error caused by the transfer for the meta-grid B (A) is $0.64 \cdot 10^{-5}$ ($0.84 \cdot 10^{-5}$) and is therefore relatively small.

When attention is paid to the additional resistance in the inertial system, shown in Figure \ref{fig:results_interpolated_vs_cfd_global}, the deviations increase. 
The pitch angle is now included in the integration step, where the local pressure fields are integrated to obtain the resistance force in global $x_1$ direction. 
Due to the time dependence of the pitch angle $\gamma(t)$, the mean of the integral and the integral of the mean values do no longer agree, and drag coefficients of both strategies deviate. The deviations increase in simulations that are associated with high pitch angle values, for example when the wave length approaches the ship length in head sea conditions i3-3 and j3-3, and decreases for low pitch angle scenarios, e.g. simulations referring to following seas (a, b). This results in a significantly higher drag error for the averaged data set strategy ($3.42 \cdot 10^{-5}$) in comparison with the integration of instantaneous pressure fields ($0.94 \cdot 10^{-5}$). 
The phenomenon becomes more pronounced and also occurs at shorter wave lengths when the angle of incidence becomes increases and the ship is exposed to  beam or head seas.

An overview of all drag coefficients determined on both meta-grids can be found in Tables \ref{tab:dragvalueslocal} and \ref{tab:dragvaluesglobal} in the appendix \ref{sec:appendixA} for the ship-fixed coordinate system and the inertia system, respectively. In the remainder of this study, only the ship-fixed coordinate system will be used to determine the quality of the entire surrogate process. The reason for this is \textendash{} as already described above \textendash{} the improved efficiency of the ML process.

\subsection{Verification of the Surrogate Model}
\label{sec:verifySurrogateModel}
The predictions of the surrogate model are compared against the CFD predictions for different model settings. The comparison includes the error of the predicted total drag magnitude in ship-fixed coordinates, i.e., $\tilde{E}_\mathrm{D} := ||C'_\mathrm{D}-\tilde{C}'_\mathrm{D,a}||_1$, as well as the RMS value of the difference between the transferred CFD pressures and the surrogate model pressure predictions 
at the meta-grid node positions, i.e., $\tilde{E}_p := \frac{||p-\tilde{p}||_2}{ ||p||_2}$.
The different settings of the ML model refer to the employed meta-grid, the handling of cells inside the ship geometry, 
as well as the training strategy. 

An overview of the results obtained for the simultaneously trained model is given in Table \ref{tab:verification}.
 In this table, the 18 test cases are ordered by increasing incidence angle (from $0$/$360\deg$ to $180\deg$), with test cases based on mirrored data sets \textendash{} i.e., $180 \deg < \beta \le 360 \deg$ \textendash{} marked with an  ``m''. 

\begin{table}[ht!]
    \caption{Overview of the absolute drag error $\tilde{E}_\mathrm{D} := ||C'_\mathrm{D}-\tilde{C}'_\mathrm{D,a}||_1$ with the true drag coefficient $C'_\mathrm{D}$ and the corresponding predicted drag coefficient $\tilde{C}'_\mathrm{D,a}$ from a simultaneous training approach as well as the relative RMS error of the pressure fields $\tilde{E}_p := \frac{||p-\tilde{p}||_2}{ ||p||_2}$ for each test case including the overall mean values. Errors are given for both meta-grids A and B as well as for both strategies to replace the NaN values of the inner ship geometry, cf. Section \ref{sec:datasets}. For the Int.-Mean (Int.-Zero) strategy all errors are represented without (with) parentheses. Outliers for values $E_\mathrm{D} > 5\cdot 10^{-5}$ or $E_p > 0.75$ are highlighted in grey.  }
    \centering
    \begin{tabular}{rcrccrrrrrrrr}
        \toprule
         \multicolumn{2}{c}{Test case} & \multicolumn{1}{c}{Inc. angle} &  \multicolumn{1}{c}{Wave length} & \multicolumn{1}{c}{Wave height} & \multicolumn{4}{c}{Abs. drag error $\tilde{E}'_\mathrm{D} / 10^{-5}$} & \multicolumn{4}{c}{Rel. RMS error $\tilde{E}_p$} \\ 
         \multicolumn{2}{c}{run}&\multicolumn{1}{c}{$\beta/$\,deg}&\multicolumn{1}{c}{$\lambda / L_\mathrm{pp}$} &\multicolumn{1}{c}{$2a/L_\mathrm{pp}$}& \multicolumn{2}{c}{Grid A}&  \multicolumn{2}{c}{Grid B}& \multicolumn{2}{c}{Grid A} &  \multicolumn{2}{c}{Grid B}
         \\  & & & &  & Mean & Zero & Mean & Zero  & Mean & Zero & Mean & Zero  
         \\\midrule
         $1$ &a2-2 & $0$     & $0.75$ & $1/100$ & $2.30$ & ($2.93$) & $1.60$ & ($3.77$) & $0.52$ & \hl{($0.92$)} & \hl{$0.78$} & \hl{($0.87$)}\\
         $2$ &b2-1 & $20$    & $0.75$ & $1/200$ & $0.23$ & ($1.14$) & $0.18$ & ($0.57$) & \hl{$1.21$}& \hl{($1.15$)} & \hl{$1.31$} & \hl{($1.00$)}\\
         $3$ &b2-3m & $340$    & $0.75$ & $1/66$ & \hl{$5.70$} & ($0.38)$ & $4.13$ & ($0.48$) & \hl{$1.25$} & \hl{($1.09)$} & \hl{$1.00$} & ($0.73$)\\
         $4$ &c3-1m & $320$    & $1.00$& $1/200$ & $1.51$ & ($0.16$) & $0.57$ & ($0.31$) & $0.49$ & ($0.63$) & $0.50$ & ($0.51$)\\
         $5$ &c4-1 & $40$    & $1.25$ & $1/200$ & $2.96$ & ($2.24$) & $1.67$ & ($1.19$) & $0.48$ & ($0.69$) & $0.54$ & ($0.54$)\\
         $6$ &d1-2m & $300$   & $0.50$ & $1/100$ & $0.30$ & ($0.55$) & $2.38$ & ($1.34$) & $0.62$ & ($0.52$) & $0.42$ & ($0.57$)\\
         $7$ &d3-2 & $60$     & $1.00$& $1/100$ & $1.78$ & ($1.70$) & $2.58$ & ($1.36$) & $0.52$ & ($0.57$) & $0.55$ & ($0.55$)\\
         $8$ &e1-3m & $280$    & $0.50$ & $1/66$ & $4.87$ & \hl{($10.80$)}& $4.65$ & \hl{($12.73$)}& \hl{$1.13$} & \hl{($1.29$)} & \hl{$1.23$} & \hl{($1.54$)}\\
         $9$ &e3-3 & $80$      & $1.00$& $1/66$ & $3.14$ & ($4.01$) & $2.12$ & ($2.91$) & $0.52$ & ($0.39$) & $0.47$ & ($0.48$)\\
         $10$&f5-1 & $100$   & $1.50$ & $1/200$ & $0.85$ & ($0.25$) & $0.30$ & ($0.75$) & \hl{$0.78$} & \hl{($0.78$)} & \hl{$0.93$} & \hl{($0.76$)}\\
         $11$&f5-3m & $260$    & $1.50$& $1/66$ & $4.04$ & ($1.99$) & $0.77$ & ($0.58$) & \hl{$0.81$} & ($0.55$) & $0.42$ & \hl{($0.89$)}\\
         $12$&g2-2 & $120$   & $0.75$& $1/100$  & $2.79$ & \hl{($6.39$)} & $3.85$ & ($3.18$) & $0.49$ & ($0.47$) & $0.29$ & ($0.25$)\\
         $13$&g3-2m & $240$  & $1.00$& $1/100$  & $5.32$ & ($3.45$) & $2.97$ & ($2.67$) & $0.42$ & ($0.52$) & $0.38$ & ($0.57$)\\
         $14$&h1-3m & $220$    & $0.50$& $1/66$ & $0.84$ & ($1.98$) & $0.95$ & ($0.08$) & $0.39$ & ($0.38$) & $0.44$ & ($0.41$)\\
         $15$&h3-1 & $140$    & $1.00$& $1/200$ & $1.88$ & ($1.45$) & $0.71$ & ($0.41$) & $0.23$ & ($0.26$) & $0.24$ & ($0.25$)\\
         $16$&i3-1m & $200$   & $1.00$& $1/200$ & $1.57$ & ($1.68$) & $0.20$ & ($2.34$) & $0.15$ & ($0.21$) & $0.25$ & ($0.25$)\\
         $17$&i5-2 & $160$    & $1.50$& $1/100$ & $1.38$ & ($1.42$) & $0.48$ & ($1.35$) & $0.17$ & ($0.15$) & $0.15$ & ($0.14$)\\
         $18$&j4-3 & $180$     & $1.25$& $1/66$ & $3.05$ & ($4.96$) & $1.80$ & \hl{($5.54$)} & $0.16$ & ($0.15$) & $0.11$ & ($0.11$)\\ \midrule 
         \multicolumn{2}{c}{Mean}&&&            & $2.47$ & ($2.63$) & $1.77$ & ($2.31$) & $0.57$ & ($0.60$) & $0.56$ & ($0.58$)\\ \midrule 
    \end{tabular}
    \label{tab:verification}
\end{table}
 In general, the Int.-Mean strategy, where the inner ship cells are filled iteratively with mean values of the boundary cells  (cf. Section \ref{sec:datasets}), shows the most promising results in combination with the better resolving meta-grid B. 
The averaged drag error $\tilde{E}_\mathrm{D}$ of all test cases is with $1.77\cdot 10^{-5}$ only twice as high as the average drag error observed in the previous chapter for the interpolation and integration step. The lower resolution meta-grid A performs slightly worse with an averaged drag error of $2.47\cdot 10^{-5}$. 

The treatment of cells inside the ships' geometry also affects the quality of the results, even though the loss function is set to zero in this part of the domain. We suspect that the convolutional layers, which take neighboring information into account, are the reason for this, as the drag is computed from cells along the boundary, i.e., cells with neighboring cells inside the ship.  Zero values (Int.-Zero) next to potentially high pressure values could have a detrimental blurring effect, whereas the Int.-Mean strategy should reconstruct these values with higher accuracy. 

The integration of forces only accounts for a small fraction of the pressure field close to the hull. Therefore, the relative RMS error $\tilde{E}_\mathrm{p}$ of the entire predicted pressure field of ML is used to evaluate the overall quality of the CAE/NN. Regarding this error, the overall picture remains unchanged. Cases with incidence angles close to $\pm 180\deg$, that is, head waves, are easier to predict, and both error measures remain relatively low. The beam-sea test case $8$ (e1-3m) represents a noticeable outlier, that performs worst for all grids and error quantities. 

\begin{figure}[ht!]
    \centering
    \includegraphics[width=0.99\textwidth]{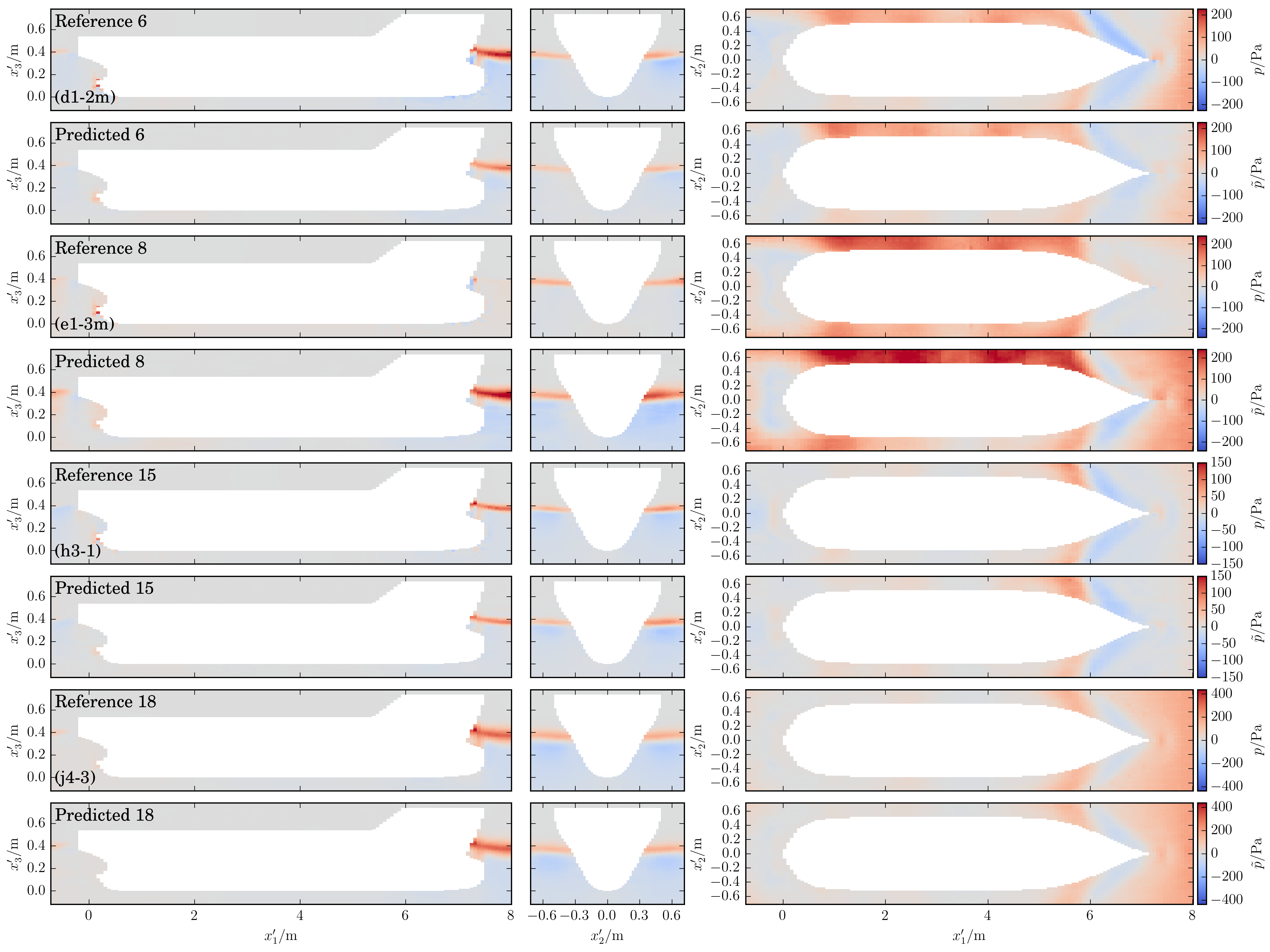}
    \caption{Comparison of pressure fields obtained from the surrogate model on meta-grid B with their corresponding CFD pressure fields for four different cases. Presentation in three different sections: $x'_1$-$x'_3$ longitudinal plane  (left), $x'_2$-$x'_3$ frame plane at $x'_1 = 6\,\si{m}$ (center) and $x'_1$-$x'_2$ top-view plane at calm-water height (right).
    For the notation used, cf. Section \ref{sec:procedure}.}
    \label{fig:results_reconstructed_pressure_comp1}  
\end{figure}

\begin{figure}[ht!]
    \centering
    \includegraphics[width=0.99\textwidth]{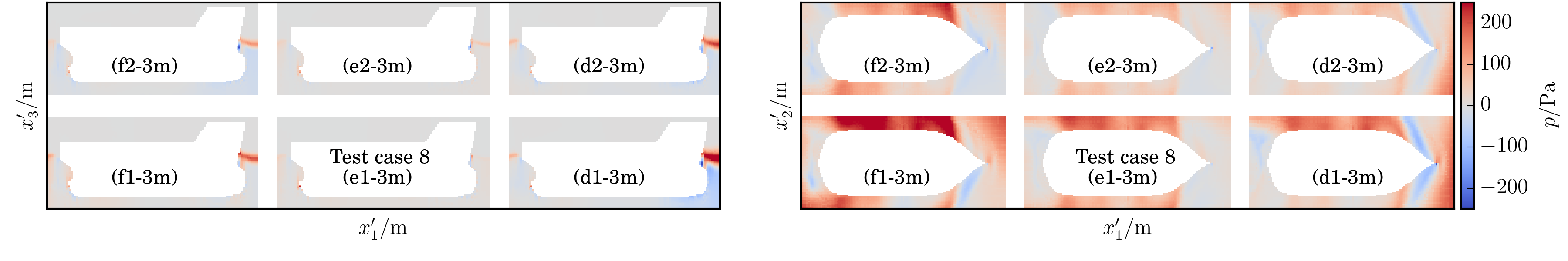}
    \caption{Comparison of transferred CFD pressure fields 
    on  meta-grid A for case 8 with neighboring wavelength and incidence angle cases, cf. Table \ref{tab:verification} for reference.}
    \label{fig:results_reconstructed_pressure_comp2}  
\end{figure}

In Figure \ref{fig:results_reconstructed_pressure_comp1}, the pressure fields predicted by the surrogate model are compared with the corresponding \emph{true} pressure fields of the CFD simulations. The figure depicts four different test cases of Table \ref{tab:verification}, i.e., case 6 (beam/following seas; d1-2m), case 8 (beam seas, e1-3m), case 15 (beam/head seas, h3-1) and case 18 (head seas; j4-3).
The pressure fields of the cases $15$ and $18$ are reproduced remarkably well, both quantitatively and qualitatively. Results for case $6$ reveal a qualitatively reasonable agreement, but deviate slightly in the bow region. Pressure fields are under predicted by the CAE/NN, which results in a slightly increased drag error of $\tilde{E}_\mathrm{D} = 2.38 \cdot 10^{-5} $ for meta-grid B. Surprisingly, the drag prediction of the coarser meta-grid A outperforms that of the finer meta-grid B, although the RMS error of the pressure field is higher for meta-grid A. This results from the generally under predicted pressure fields in the bow and stern areas, and highlights the importance of considering multiple error metrics. In contrast, the pressure field of test case $8$ is massively over predicted by the surrogate model. This can be explained by the size of the underlying data set and the corresponding pressure patterns for other simulations with similar wave length and incidence angle, cf. Figure \ref{fig:results_reconstructed_pressure_comp2}. The pressure patterns of the neighboring cases f1-3m and d1-3m significantly deviate from the pressure field of case $8$. Therefore, the dataset size appears to be too small to represent the pressure fields with uniform accuracy for the entire parameter space. 

\subsection{Hyperparameter and Training Influences}
\label{sec:hyperTraining}
The model performance is also analyzed for a separate training approach. Without presenting details, it can be stated that errors did slightly increase on average for the separate training, resulting in a relative RMS error of $0.59$ in combination with meta-grid B and the Int.-Mean strategy, i.e. approximately 5\% more than the value ($0.56$, cf. Table \ref{tab:verification}) for a simultaneous training approach. 
On the other hand, outliers seem to be slightly less emphasized for the separate training. Since the simultaneous training approach also performs best for this grid in terms of drag, it remains the better choice for this study.

Finally, the influence of the dimension of the latent space on the quality of the predictions is evaluated. In addition to the default latent space dimension of $10$ (cf. Table \ref{tab:regression_net}), two further trainings are performed for a halved and doubled latent space, resulting in a dimension of $5$ and $20$, respectively. In terms of the averaged RMS error of the pressure fields, both variations perform slightly worse, resulting in an error of $0.59$ ($0.65$) for a lower (higher) latent dimension of $5$ ($20$). The default latent space dimension of $10$ seems to be a good compromise between capturing all the necessary features of the pressure fields that occur and the training effort given by the size of the data set.

\section{Conclusion and Outlook}
\label{sec:conclusion}
This paper presents an autoencoder-based surrogate modeling approach for the prediction of pressure fields and related added resistance forces on ships in  sea-states. To this end, $150$ different sea-state conditions are simulated using a finite-volume CFD approach, which serve as the data basis for the surrogate modeling process chain. Various surrogate model settings are evaluated, including two different meta-grids, two strategies for managing cells within the ship geometry, as well as different hyperparameter and training options. 

Satisfactory predictive accuracy was achieved for both, the  pressure fields as well as the related drag coefficients for the majority of test cases. The pressure predictions of the surrogate model are superior for cases with incidence angles close to $\pm 180\deg$ (head seas). Observed deviations in some test cases might benefit from a locally larger training set, especially for following seas, where more complex pressure pattern and flow conditions occur.
As expected, the finer meta-grid provides superior results, particularly in combination with the iterative averaging of data in the interior of the ship geometry (Int.-Mean). 

A surrogate model of this type could be used in an onboard decision support system to rapidly predict additional resistance forces for different wave scenarios prevailing at sea.
This work could also point the way for AI-supported surrogate modeling of complex multiphase flow fields. Existing CFD data sets could serve as a database for future CAE/NN surrogate models.
Future work may focus on the processing of bigger and potentially unstructured data sets, improved training strategies that might separate relevant from less relevant latent variables, as well as the application to further engineering problems.

\section*{Acknowledgments}
The authors gratefully acknowledge the computing time made available to them on
the high-performance computers Lise and Emmy/Grete at the NHR Centers ZIB and GWDG. These
Centers are jointly supported by the Federal Ministry of Education and Research
and the state governments participating in the NHR (www.nhr-verein.de/unserepartner).

This work is a part of the research projects \emph{Near real-time modeling and simulation of environmental influences on ship energy consumption} funded by the German Federal Ministry for Economic Affairs and Energy [BMWi, grant number 03SX528L]. 
We acknowledge support for the Open Access fees by Hamburg University of Technology (TUHH) in the funding program Open Access Publishing.

\section*{Disclosure of interest}
The authors report there are no competing interests to declare.

\section*{Data availability statement}
The training data that support the findings of this study are openly available in TUHH Open Research (TORE) at https://doi.org/10.15480/882.15084 after publication. Raw CFD data are available from the corresponding author upon reasonable request.


\appendix
\section{Drag coefficients}
\label{sec:appendixA}
An overview of all simulated and data processed drag coefficients is given in the Tables \ref{tab:dragvalueslocal} and \ref{tab:dragvaluesglobal} below. 
\begin{table}[ht!]
    \addtolength{\tabcolsep}{-3pt} 
    \caption{Overview of all local drag coefficients based on CFD simulations ($\Delta C'_\mathrm{D}$) as well as reconstructed values from time averaged ($\Delta C'_\mathrm{D,a}$) and instantaneous flow data ($\Delta C'_\mathrm{D,i}$) for two meta-grids A and B. The mean drag error $E'_\mathrm{D} := ||C'_\mathrm{D}-C'_\mathrm{D,a,i}||_1$ is given at the end of the table.} 
    \centering
    \begin{tabular}{rrrrrr rrrrrr rrrrrr}
        \toprule
        \multicolumn{1}{c}{\rotatebox[origin=c]{90}{Run}} & \multicolumn{1}{c}{\rotatebox[origin=c]{90}{$\Delta C'_\mathrm{D}/ 10^{-5}$}} & \multicolumn{1}{c}{\rotatebox[origin=c]{90}{$\Delta C'_\mathrm{D,a}/ 10^{-5}$}}&  \multicolumn{1}{c}{\rotatebox[origin=c]{90}{$\Delta C'_\mathrm{D,i}/ 10^{-5}$}} &  \multicolumn{1}{c}{\rotatebox[origin=c]{90}{$\Delta C'_\mathrm{D,a} /10^{-5}$}} &  \multicolumn{1}{c}{\rotatebox[origin=c]{90}{$\Delta C'_\mathrm{D,i} /10^{-5}$}} & \multicolumn{1}{c}{\rotatebox[origin=c]{90}{Run}} & \multicolumn{1}{c}{\rotatebox[origin=c]{90}{$\Delta C'_\mathrm{D}/ 10^{-5}$}} & \multicolumn{1}{c}{\rotatebox[origin=c]{90}{$\Delta C'_\mathrm{D,a}/ 10^{-5}$}}&  \multicolumn{1}{c}{\rotatebox[origin=c]{90}{$\Delta C'_\mathrm{D,i}/ 10^{-5}$}} &  \multicolumn{1}{c}{\rotatebox[origin=c]{90}{$\Delta C'_\mathrm{D,a} /10^{-5}$}}&  \multicolumn{1}{c}{\rotatebox[origin=c]{90}{$\Delta C'_\mathrm{D,i}/ 10^{-5}$}} & \multicolumn{1}{c}{\rotatebox[origin=c]{90}{Run}} & \multicolumn{1}{c}{\rotatebox[origin=c]{90}{$\Delta C'_\mathrm{D}/ 10^{-5}$}} & \multicolumn{1}{c}{\rotatebox[origin=c]{90}{$\Delta C'_\mathrm{D,a} /10^{-5}$}}&  \multicolumn{1}{c}{\rotatebox[origin=c]{90}{$\Delta C'_\mathrm{D,i} /10^{-5}$}} &  \multicolumn{1}{c}{\rotatebox[origin=c]{90}{$\Delta C'_\mathrm{D,a}/ 10^{-5}$}} &  \multicolumn{1}{c}{\rotatebox[origin=c]{90}{$\Delta C'_\mathrm{D,i}/ 10^{-5}$}} \\ \cmidrule(lr){3-4} \cmidrule(lr){5-6} \cmidrule(lr){9-10} \cmidrule(lr){11-12} \cmidrule(lr){15-16} \cmidrule(lr){17-18}
        &&\multicolumn{2}{c}{Grid A} & \multicolumn{2}{c}{Grid B} & & & \multicolumn{2}{c}{Grid A} & \multicolumn{2}{c}{Grid B} & & & \multicolumn{2}{c}{Grid A} & \multicolumn{2}{c}{Grid B} \\ \midrule 
        & & & & & & \multicolumn{1}{c}{\vdots} & \multicolumn{1}{c}{\vdots} & \multicolumn{1}{c}{\vdots} & \multicolumn{1}{c}{\vdots} & \multicolumn{1}{c}{\vdots} & \multicolumn{1}{c}{\vdots} & \multicolumn{1}{c}{\vdots} & \multicolumn{1}{c}{\vdots} & \multicolumn{1}{c}{\vdots} & \multicolumn{1}{c}{\vdots} & \multicolumn{1}{c}{\vdots} & \multicolumn{1}{c}{\vdots} \\
        a1-1 &$8.92$ & $6.54$ &$6.54$ & $7.40$&$7.40$ & d3-1 &$2.14$&  $1.36$& $1.36$& $1.89$& $1.89$& g5-1 &$1.70$&  $1.28$& $1.28$& $1.63$& $1.63$ \\
        a1-2 &$2.08$ & $0.00$&$0.00$&$0.87$ &$0.87$ & d3-2 &$3.72$&  $3.84$& $3.84$& $4.50$& $4.50$& g5-2 &$3.85$&  $4.08$& $4.08$& $4.25$& $4.25$ \\
        a1-3 &$-5.61$&$-7.69$&$-7.69$&$-6.62$&$-6.62$& d3-3 &$3.54$&  $3.44$& $3.44$& $3.98$& $3.98$& g5-3 &$9.15$&  $10.69$& $10.69$& $10.80$& $10.80$\\
        a2-1 &$3.21$&$2.76$&$2.76$&$3.34$&$3.34$& d4-1 &$1.91$&  $0.81$& $0.81$& $1.38$& $1.38$& h1-1 &$5.27$&  $3.55$& $3.55$& $4.64$& $4.64$ \\
        a2-2 &$0.63$&$0.30$&$0.30$&$0.87$&$0.87$& d4-2 &$3.39$&  $2.30$& $2.30$& $3.05$& $3.05$& h1-2 &$6.70$&  $5.45$& $5.45$& $6.44$& $6.44$ \\
        a2-3 &$-1.89$&$-2.03$&$-2.03$&$-1.47$&$-1.47$& d4-3 &$2.71$&  $2.08$& $2.08$& $2.69$& $2.69$& h1-3 &$10.22$&  $9.37$& $9.37$& $10.34$& $10.34$  \\
        a3-1 &$1.94$&$1.74$&$1.74$&$2.10$&$2.10$& d5-1 &$1.72$&  $0.53$& $0.53$& $1.03$& $1.03$& h2-1 &$4.53$&  $4.74$& $4.74$& $5.05$& $5.05$ \\
        a3-2 &$2.80$&$2.78$&$2.78$&$2.96$&$2.96$& d5-2 &$2.73$&  $1.94$& $1.94$& $2.92$& $2.92$& h2-2 &$8.72$&  $9.81$& $9.81$& $10.01$& $10.01$ \\
        a3-3 &$3.59$&  $6.00$& $6.00$& $5.05$& $5.05$& d5-3 &$2.85$&  $2.63$& $2.63$& $3.47$& $3.47$& h2-3 &$16.68$&  $18.95$& $18.95$& $18.71$& $18.71$ \\
        a4-1 &$1.41$&  $0.98$& $0.98$& $1.41$& $1.41$& e1-1 &$2.96$&  $1.31$& $1.31$& $2.39$& $2.39$& h3-1 &$1.68$&  $1.99$& $1.99$& $2.26$& $2.26$\\
        a4-2 &$1.77$&  $2.04$& $2.04$& $2.40$& $2.40$& e1-2 &$4.66$&  $2.97$& $2.97$& $4.01$& $4.01$& h3-2 &$5.86$&  $7.19$& $7.19$& $7.33$& $7.33$ \\
        a4-3 &$2.62$&  $3.11$& $3.11$& $3.40$& $3.40$& e1-3 &$8.12$&  $6.74$& $6.74$& $7.77$& $7.77$& h3-3 &$16.68$&  $19.43$& $19.43$& $19.30$& $19.30$ \\
        a5-1 &$0.92$&  $0.42$& $0.42$& $0.81$& $0.81$& e2-1 &$3.50$&  $2.67$& $2.67$& $3.30$& $3.30$& h4-1 &$0.90$&  $0.54$& $0.54$& $1.00$& $1.00$ \\
        a5-2 &$0.43$&  $0.33$& $0.33$& $0.68$& $0.68$& e2-2 &$4.73$&  $4.90$& $4.90$& $5.60$& $5.60$& h4-2 &$1.88$&  $2.19$& $2.19$& $2.51$& $2.51$ \\
        a5-3 &$1.28$&  $1.78$& $1.78$& $2.09$& $2.09$& e2-3 &$6.70$&  $6.60$& $6.60$& $7.23$& $7.23$& h4-3 &$4.23$&  $5.60$& $5.60$& $5.83$& $5.83$ \\
        b1-1 &$-5.99$&  $-7.36$& $-7.36$& $-6.06$& $-6.06$& e3-1 &$1.22$&  $0.66$& $0.66$& $1.19$& $1.19$& h5-1 &$1.66$&  $1.15$& $1.15$& $1.48$& $1.48$ \\
        b1-2 &$-6.62$&  $-7.93$& $-7.93$& $-6.73$& $-6.73$& e3-2 &$2.02$&  $2.58$& $2.58$& $3.31$& $3.31$& h5-2 &$4.14$&  $4.08$& $4.08$& $4.24$& $4.24$ \\
        b1-3 &$-7.18$&  $-8.41$& $-8.41$& $-7.26$& $-7.26$& e3-3 &$2.41$&  $2.39$& $2.39$& $2.88$& $2.88$& h5-3 &$8.37$&  $9.20$& $9.20$& $9.00$& $9.00$ \\
        b2-1 &$1.40$&  $1.09$& $1.09$& $1.65$& $1.65$& e4-1 &$1.95$&  $0.82$& $0.82$& $1.39$& $1.39$& i1-1 &$3.13$&  $1.50$& $1.50$& $2.68$& $2.68$ \\
        b2-2 &$2.22$&  $2.14$& $2.14$& $2.57$& $2.57$& e4-2 &$3.27$&  $2.36$& $2.36$& $3.18$& $3.18$& i1-2 &$4.55$&  $3.16$& $3.16$& $4.28$& $4.28$ \\
        b2-3 &$2.94$&  $3.17$& $3.17$& $3.51$& $3.51$& e4-3 &$2.30$&  $1.60$& $1.60$& $2.10$& $2.10$& i1-3 &$7.01$&  $5.76$& $5.76$& $6.83$& $6.83$ \\
        b3-1 &$2.07$&  $1.90$& $1.90$& $2.26$& $2.26$& e5-1 &$1.61$&  $0.68$& $0.68$& $1.10$& $1.10$& i2-1 &$2.80$&  $2.66$& $2.66$& $3.15$& $3.15$ \\
        b3-2 &$2.14$&  $2.30$& $2.30$& $2.52$& $2.52$& e5-2 &$2.40$&  $2.48$& $2.48$& $3.17$& $3.17$& i2-2 &$4.70$&  $4.80$& $4.80$& $5.14$& $5.14$ \\
        b3-3 &$1.68$&  $2.19$& $2.19$& $2.42$& $2.42$& e5-3 &$2.46$&  $1.79$& $1.79$& $2.31$& $2.31$& i2-3 &$8.24$&  $8.92$& $8.92$& $9.17$& $9.17$ \\
        b4-1 &$1.51$&  $0.74$& $0.74$& $1.33$& $1.33$& f1-1 &$4.38$&  $3.03$& $3.03$& $4.18$& $4.18$& i3-1 &$4.62$&  $4.75$& $4.75$& $4.91$& $4.91$ \\
        b4-2 &$1.37$&  $0.76$& $0.76$& $1.31$& $1.31$& f1-2 &$10.52$&  $9.60$& $9.60$& $10.56$& $10.56$& i3-2 &$18.42$&  $19.81$& $19.81$& $19.51$& $19.51$\\
        b4-3 &$0.97$&  $0.80$& $0.80$& $1.37$& $1.37$& f1-3 &$22.84$&  $21.92$& $21.92$& $22.73$& $22.73$& i3-3 &$48.89$&  $53.89$& $53.89$& $53.15$& $53.15$ \\
        b5-1 &$1.15$&  $0.26$& $0.26$& $0.39$& $0.39$& f2-1 &$2.59$&  $2.42$& $2.42$& $2.88$& $2.88$& i4-1 &$1.16$&  $0.94$& $0.94$& $1.33$& $1.33$\\
        b5-2 &$1.55$&  $0.85$& $0.85$& $1.24$& $1.24$& f2-2 &$3.89$&  $3.86$& $3.86$& $4.28$& $4.28$& i4-2 &$2.90$&  $3.44$& $3.44$& $3.86$& $3.86$ \\
        b5-3 &$1.42$&  $0.94$& $0.94$& $1.47$& $1.47$& f2-3 &$4.67$&  $5.23$& $5.23$& $5.81$& $5.81$& i4-3 &$7.43$&  $9.09$& $9.09$& $9.14$& $9.14$ \\
        c1-1 &$5.26$&  $3.61$& $3.61$& $4.62$& $4.62$& f3-1 &$1.27$&  $0.99$& $0.99$& $1.39$& $1.39$& i5-1 &$1.59$&  $-0.01$& $-0.01$& $0.04$& $0.04$ \\
        c1-2 &$8.21$&  $6.94$& $6.94$& $7.75$& $7.765$& f3-2 &$3.73$&  $3.73$& $3.73$& $4.160$& $4.16$& i5-2 &$4.02$&  $4.32$& $4.32$& $4.51$& $4.51$  \\
        c1-3 &$11.44$&  $10.65$& $10.65$& $11.30$& $11.30$& f3-3 &$5.95$&  $6.57$& $6.57$& $6.91$& $6.91$& i5-3 &$8.05$&  $9.18$& $9.18$& $9.30$& $9.30$ \\
        c2-1 &$-0.69$&  $-0.59$& $-0.59$& $-0.16$& $-0.16$& f4-1 &$2.02$&  $1.25$& $1.25$& $1.70$& $1.70$& j1-1 &$3.49$&  $1.89$& $1.89$& $3.02$& $3.02$ \\
        c2-2 &$-2.84$&  $-1.95$& $-1.95$& $-1.63$& $-1.63$& f4-2 &$3.93$&  $3.28$& $3.28$& $3.76$& $3.76$& j1-2 &$5.64$&  $4.59$& $4.59$& $5.62$& $5.62$ \\
        c2-3 &$-4.32$&  $-2.29$& $-2.29$& $-2.06$& $-2.06$& f4-3 &$5.73$&  $5.59$& $5.59$& $5.93$& $5.93$& j1-3 &$10.10$&  $9.15$& $9.15$& $10.15$& $10.15$ \\
        c3-1 &$1.32$&  $0.98$& $0.98$& $1.38$& $1.38$& f5-1 &$1.84$&  $1.07$& $1.07$& $1.39$& $1.39$& j2-1 &$2.99$&  $2.88$& $2.88$& $3.36$& $3.36$ \\
        c3-2 &$1.88$&  $2.02$& $2.02$& $2.49$& $2.49$& f5-2 &$3.56$&  $2.77$& $2.77$& $3.22$& $3.22$& j2-2 &$4.91$&  $5.03$& $5.03$& $5.35$& $5.35$ \\
        c3-3 &$2.67$&  $4.09$& $4.09$& $4.58$& $4.58$& f5-3 &$5.37$&  $5.16$& $5.16$& $5.43$& $5.43$& j2-3 &$8.46$&  $9.21$& $9.21$& $9.48$& $9.48$ \\
        c4-1 &$1.12$&  $0.33$& $0.33$& $0.87$& $0.87$& g1-1 &$2.71$&  $1.27$& $1.27$& $2.37$& $2.37$& j3-1 &$2.47$&  $3.21$& $3.21$& $3.37$& $3.37$ \\
        c4-2 &$1.21$&  $0.81$& $0.81$& $1.42$& $1.42$& g1-2 &$4.79$&  $4.36$& $4.36$& $5.16$& $5.16$& j3-2 &$13.37$&  $15.02$& $15.02$& $14.64$& $14.64$ \\
        c4-3 &$3.07$&  $4.00$& $4.00$& $4.59$& $4.59$& g1-3 &$9.44$&  $9.44$& $9.44$& $10.17$& $10.17$& j3-3 &$44.93$&  $49.99$& $49.99$& $49.32$& $49.32$ \\
        c5-1 &$1.33$&  $0.80$& $0.80$& $1.21$& $1.21$& g2-1 &$5.91$&  $6.38$& $6.38$& $6.68$& $6.68$& j4-1 &$2.87$&  $2.47$& $2.467$& $2.79$& $2.79$ \\
        c5-2 &$2.36$&  $2.82$& $2.82$& $3.14$& $3.14$& g2-2 &$14.66$&  $16.36$& $16.36$& $16.34$& $16.34$& j4-2 &$6.28$&  $6.49$& $6.49$& $6.89$& $6.89$ \\
        c5-3 &$8.19$&  $9.54$& $9.54$& $9.86$& $9.86$& g2-3 &$33.57$&  $37.44$& $37.44$& $37.22$& $37.22$& j4-3 &$13.06$&  $14.23$& $14.23$& $14.14$& $14.14$ \\
        d1-1 &$2.65$&  $1.00$& $1.00$& $2.13$& $2.13$& g3-1 &$1.17$&  $1.10$& $1.10$& $1.42$& $1.42$& j5-1 &$2.37$&  $1.94$& $1.94$& $2.27$& $2.27$ \\
        d1-2 &$5.17$&  $4.22$& $4.22$& $5.32$& $5.32$& g3-2 &$3.24$&  $3.75$& $3.75$& $3.94$& $3.94$& j5-2 &$5.07$&  $4.97$& $4.97$& $5.21$& $5.21$ \\
        d1-3 &$9.91$&  $10.39$& $10.39$& $11.32$& $11.32$& g3-3 &$6.71$&  $8.02$& $8.02$& $8.14$& $8.14$& j5-3 &$9.40$&  $9.94$& $9.94$& $10.03$& $10.03$ \\ 
        d2-1 &$3.86$&  $3.37$& $3.37$& $3.92$& $3.92$& g4-1 &$2.38$&  $1.75$& $1.75$& $2.18$& $2.18$& & & & & & \\
        d2-2 &$5.62$&  $5.88$& $5.88$& $6.41$& $6.41$& g4-2 &$4.12$&  $3.99$& $3.99$& $4.30$& $4.30$& & & & & & \\
        d2-3 &$8.45$&  $9.73$& $9.73$& $10.21$& $10.21$& g4-3 &$7.25$&  $7.31$& $7.31$& $7.47$& $7.47$& & & & & & \\\cmidrule(lr){13-18}
        \multicolumn{1}{c}{\vdots} & \multicolumn{1}{c}{\vdots} & \multicolumn{1}{c}{\vdots} & \multicolumn{1}{c}{\vdots} & \multicolumn{1}{c}{\vdots} & \multicolumn{1}{c}{\vdots} & \multicolumn{1}{c}{\vdots} & \multicolumn{1}{c}{\vdots} & \multicolumn{1}{c}{\vdots} & \multicolumn{1}{c}{\vdots} & \multicolumn{1}{c}{\vdots} & \multicolumn{1}{c}{\vdots} & \multicolumn{2}{c}{$E'_\mathrm{D} / 10^{-5}$} &$0.84$ & $0.84$ & $0.65$& $0.65$ \\ 
        \bottomrule
   
    \end{tabular}
    
    \label{tab:dragvalueslocal}
\end{table}

\begin{table}[ht!]
    \addtolength{\tabcolsep}{-3.5pt} 
    \caption{Overview of all global drag coefficients based on CFD simulations ($\Delta C_\mathrm{D}$) as well as reconstructed values from time averaged ($\Delta C_\mathrm{D,a}$) and instantaneous flow data ($\Delta C_\mathrm{D,i}$) for two meta-grids A and B. The mean drag error $E_\mathrm{D} := ||C_\mathrm{D}-C_\mathrm{D,a,i}||_1$ is given at the end of the table.} 
    \centering
    \begin{tabular}{rrrrrr rrrrrr rrrrrr}
        \toprule
        \multicolumn{1}{c}{\rotatebox[origin=c]{90}{Run}} & \multicolumn{1}{c}{\rotatebox[origin=c]{90}{$\Delta C_\mathrm{D}/ 10^{-5}$}} & \multicolumn{1}{c}{\rotatebox[origin=c]{90}{$\Delta C_\mathrm{D,a}/ 10^{-5}$}}&  \multicolumn{1}{c}{\rotatebox[origin=c]{90}{$\Delta C_\mathrm{D,i}/ 10^{-5}$}} &  \multicolumn{1}{c}{\rotatebox[origin=c]{90}{$\Delta C_\mathrm{D,a}/ 10^{-5}$}} &  \multicolumn{1}{c}{\rotatebox[origin=c]{90}{$\Delta C_\mathrm{D,i}/ 10^{-5}$}} & \multicolumn{1}{c}{\rotatebox[origin=c]{90}{Run}} & \multicolumn{1}{c}{\rotatebox[origin=c]{90}{$\Delta C_\mathrm{D}/ 10^{-5}$}} & \multicolumn{1}{c}{\rotatebox[origin=c]{90}{$\Delta C_\mathrm{D,a}/ 10^{-5}$}}&  \multicolumn{1}{c}{\rotatebox[origin=c]{90}{$\Delta C_\mathrm{D,i}/ 10^{-5}$}} &  \multicolumn{1}{c}{\rotatebox[origin=c]{90}{$\Delta C_\mathrm{D,a}/ 10^{-5}$}}&  \multicolumn{1}{c}{\rotatebox[origin=c]{90}{$\Delta C_\mathrm{D,i}/ 10^{-5}$}} & \multicolumn{1}{c}{\rotatebox[origin=c]{90}{Run}} & \multicolumn{1}{c}{\rotatebox[origin=c]{90}{$\Delta C_\mathrm{D}/ 10^{-5}$}} & \multicolumn{1}{c}{\rotatebox[origin=c]{90}{$\Delta C_\mathrm{D,a}/ 10^{-5}$}}&  \multicolumn{1}{c}{\rotatebox[origin=c]{90}{$\Delta C_\mathrm{D,i}/ 10^{-5}$}} &  \multicolumn{1}{c}{\rotatebox[origin=c]{90}{$\Delta C_\mathrm{D,a}/ 10^{-5}$}} &  \multicolumn{1}{c}{\rotatebox[origin=c]{90}{$\Delta C_\mathrm{D,i}/ 10^{-5}$}} \\ \cmidrule(lr){3-4} \cmidrule(lr){5-6} \cmidrule(lr){9-10} \cmidrule(lr){11-12} \cmidrule(lr){15-16} \cmidrule(lr){17-18}
        &&\multicolumn{2}{c}{Grid A} & \multicolumn{2}{c}{Grid B} & & & \multicolumn{2}{c}{Grid A} & \multicolumn{2}{c}{Grid B} & & & \multicolumn{2}{c}{Grid A} & \multicolumn{2}{c}{Grid B} \\ \midrule 
        & & & & & & \multicolumn{1}{c}{\vdots} & \multicolumn{1}{c}{\vdots} & \multicolumn{1}{c}{\vdots} & \multicolumn{1}{c}{\vdots} & \multicolumn{1}{c}{\vdots} & \multicolumn{1}{c}{\vdots} & \multicolumn{1}{c}{\vdots} & \multicolumn{1}{c}{\vdots} & \multicolumn{1}{c}{\vdots} & \multicolumn{1}{c}{\vdots} & \multicolumn{1}{c}{\vdots} & \multicolumn{1}{c}{\vdots} \\
        a1-1 &$6.89$&  $5.34$& $5.30$& $6.22$& $6.18$& d3-1 &$-0.53$&  $-0.39$& $-0.96$& $0.16$& $-0.42$& g5-1 &$2.19$&  $2.55$& $1.55$& $2.89$& $1.89$ \\
        a1-2 &$7.08$&  $6.05$& $6.00$& $6.89$& $6.84$& d3-2 &$1.49$&  $4.33$& $1.94$& $4.98$& $2.60$& g5-2 &$5.61$&  $9.58$& $5.44$& $9.72$& $5.61$ \\
        a1-3 &$4.46$&  $3.57$& $3.51$& $4.58$& $4.52$& d3-3 &$-1.22$&  $4.36$& $-0.74$& $4.90$& $-0.18$& g5-3 &$13.95$&  $24.65$& $15.06$& $24.69$& $15.13$\\
        a2-1 &$2.40$&  $2.24$& $2.22$& $2.82$& $2.80$& d4-1 &$2.29$&  $1.59$& $1.21$& $2.16$& $1.78$& h1-1 &$5.50$&  $4.73$& $4.73$& $5.81$& $5.81$ \\
        a2-2 &$-3.04$&  $-3.07$& $-3.11$& $-2.48$& $-2.52$& d4-2 &$5.59$&  $5.85$& $4.32$& $6.57$& $5.05$& h1-2 &$10.92$&  $10.64$& $10.62$& $11.61$& $11.58$ \\
        a2-3 &$-6.21$&  $-6.35$& $-6.40$& $-5.76$& $-5.82$& d4-3&$5.30$&  $7.99$& $4.75$& $8.57$& $5.34$& h1-3 &$19.14$&  $19.46$& $19.39$& $20.38$& $20.31$  \\
        a3-1 &$-0.88$&  $-1.14$& $-1.08$& $-0.77$& $-0.71$& d5-1 &$3.59$&  $2.71$& $2.53$& $3.20$& $3.02$& h2-1 &$3.53$&  $4.79$& $3.98$& $5.10$& $4.31$ \\
        a3-2 &$-3.70$&  $-4.20$& $-3.97$& $-3.98$& $-3.74$& d5-2 &$9.44$&  $9.75$& $9.05$& $10.69$& $9.99$& h2-2 &$11.60$&  $16.31$& $12.97$& $16.48$& $13.18$ \\
        a3-3 &$-6.38$&  $-5.23$& $-4.71$& $-6.12$& $-5.57$& d5-3 &$11.07$&  $12.87$& $11.36$& $13.66$& $12.14$& h2-3 &$25.75$&  $35.95$& $28.43$& $35.63$& $28.20$ \\
        a4-1 &$-4.98$&  $-5.82$& $-5.77$& $-5.35$& $-5.30$& e1-1 &$2.38$&  $1.28$& $1.54$& $2.36$& $2.63$& h3-1 &$3.49$&  $5.74$& $4.09$& $6.00$& $4.37$\\
        a4-2 &$-4.74$&  $-11.14$& $-10.94$& $-10.71$& $-10.50$& e1-2 &$6.78$&  $5.25$& $6.14$& $6.29$& $7.18$& h3-2 &$11.29$&  $19.56$& $13.07$& $19.65$& $13.21$ \\
        a4-3 &$-14.24$&  $-16.46$& $-16.10$& $-16.06$& $-15.70$& e1-3 &$11.15$&  $8.71$& $10.93$& $9.74$& $11.95$& h3-3 &$26.22$&  $43.74$& $29.66$& $43.49$& $29.50$ \\
        a5-1 &$4.76$&  $4.58$& $4.62$& $4.95$& $4.98$& e2-1 &$0.19$&  $-0.25$& $-0.51$& $0.39$& $0.13$& h4-1 &$2.49$&  $2.81$& $1.98$& $3.26$& $2.43$ \\
        a5-2 &$8.39$&  $8.21$& $8.32$& $8.52$& $8.63$& e2-2 &$1.38$&  $2.91$& $1.90$& $3.62$& $2.62$& h4-2&$6.12$&  $9.52$& $6.17$& $9.80$& $6.47$ \\
        a5-3 &$15.14$&  $15.70$& $15.92$& $15.94$& $16.16$& e2-3 &$-1.25$&  $0.93$& $-1.12$& $1.60$& $-0.43$& h4-3 &$11.74$&  $20.21$& $12.81$& $20.36$& $13.01$ \\
        b1-1 &$-4.85$&  $-5.30$& $-5.33$& $-4.01$& $-4.04$& e3-1 &$1.79$&  $2.39$& $1.47$& $2.91$& $2.01$& h5-1 &$0.96$&  $1.52$& $0.71$& $1.85$& $1.05$ \\
        b1-2 &$-3.31$&  $-3.78$& $-3.84$& $-2.61$& $-2.67$& e3-2 &$3.97$&  $8.03$& $4.48$& $8.73$& $5.21$& h5-2 &$3.77$&  $7.50$& $4.29$& $7.65$& $4.46$ \\
        b1-3 &$-0.69$&  $-0.77$& $-0.86$& $0.34$& $0.25$& e3-3 &$1.20$&  $9.69$& $1.89$& $10.14$& $2.40$& h5-3 &$10.10$&  $19.39$& $12.80$& $19.14$& $12.60$ \\
        b2-1 &$2.34$&  $2.34$& $2.34$& $2.89$& $2.90$& e4-1 &$1.80$&  $1.25$& $0.77$& $1.82$& $1.33$& i1-1 &$3.29$&  $2.52$& $2.51$& $3.70$& $3.69$ \\
        b2-2 &$4.35$&  $4.60$& $4.62$& $5.02$& $5.05$& e4-2 &$3.98$&  $4.94$& $2.98$& $5.74$& $3.81$& i1-2 &$8.09$&  $7.67$& $7.65$& $8.77$& $8.75$ \\
        b2-3 &$6.65$&  $7.21$& $7.27$& $7.53$& $7.59$& e4-3 &$1.17$&  $4.58$& $0.45$& $5.07$& $0.96$& i1-3&$15.60$&  $15.59$& $15.52$& $16.61$& $16.53$ \\
        b3-1 &$-2.74$&  $-2.82$& $-2.79$& $-2.43$& $-2.40$& e5-1 &$1.27$&  $0.53$& $0.28$& $0.95$& $0.71$& i2-1 &$2.02$&  $2.34$& $2.11$& $2.84$& $2.62$ \\
        b3-2 &$-5.99$&  $-5.94$& $-5.84$& $-5.67$& $-5.57$& e5-2 &$4.33$&  $5.55$& $4.50$& $6.22$& $5.18$& i2-2 &$6.58$&  $8.11$& $7.09$& $8.44$& $7.43$ \\
        b3-3 &$-9.84$&  $-8.93$& $-8.72$& $-8.64$& $-8.42$& e5-3 &$2.15$&  $3.67$& $1.49$& $4.18$& $2.02$& i2-3 &$14.78$&  $18.35$& $16.01$& $18.56$& $16.26$ \\
        b4-1 &$3.99$&  $3.33$& $3.33$& $3.91$& $3.90$& f1-1 &$8.05$&  $6.17$& $7.41$& $7.31$& $8.53$& i3-1 &$6.12$&  $10.94$& $7.31$& $11.08$& $7.49$ \\
        b4-2 &$6.64$&  $6.24$& $6.21$& $6.76$& $6.73$& f1-2 &$19.67$&  $14.64$& $20.09$& $15.59$& $20.95$& i3-2 &$24.51$&  $41.39$& $27.20$& $40.98$& $26.93$\\
        b4-3 &$8.81$&  $9.47$& $9.37$& $9.99$& $9.89$& f1-3 &$38.30$&  $25.56$& $39.23$& $26.37$& $39.81$& i3-3 &$55.79$&  $93.96$& $62.57$& $93.02$& $61.89$ \\
        b5-1 &$0.60$&  $-0.26$& $-0.15$& $-0.11$& $-0.01$& f2-1 &$2.01$&  $1.61$& $2.06$& $2.08$& $2.53$& i4-1 &$2.42$&  $3.17$& $2.01$& $3.56$& $2.40$\\
        b5-2 &$2.20$&  $1.55$& $1.51$& $1.94$& $1.89$& f2-2 &$3.30$&  $2.81$& $3.97$& $3.24$& $4.38$& i4-2 &$7.05$&  $11.88$& $7.25$& $12.26$& $7.68$ \\
        b5-3 &$4.26$&  $3.95$& $3.84$& $4.47$& $4.34$& f2-3 &$6.70$&  $6.43$& $8.86$& $7.01$& $9.40$& i4-3 &$15.40$&  $26.87$& $16.64$& $26.83$& $16.70$ \\
        c1-1 &$3.34$&  $2.48$& $2.46$& $3.49$& $3.48$& f3-1 &$-0.30$&  $0.19$& $-0.53$& $0.60$& $-0.12$& i5-1 &$0.94$&  $0.57$& $-0.42$& $0.62$& $-0.36$ \\
        c1-2 &$7.04$&  $6.61$& $6.60$& $7.43$& $7.42$& f3-2 &$1.58$&  $4.72$& $1.78$& $5.14$& $2.21$& i5-2 &$3.80$&  $6.76$& $3.64$& $6.95$& $3.85$ \\
        c1-3 &$12.70$&  $12.85$& $12.85$& $13.50$& $13.50$& f3-3 &$2.43$&  $9.72$& $3.25$& $10.05$& $3.60$& i5-3 &$8.98$&  $16.49$& $9.49$& $16.57$& $9.62$ \\
        c2-1 &$-7.33$&  $-7.51$& $-7.74$& $-7.04$& $-7.28$& f4-1 &$0.89$&  $0.78$& $-0.08$& $1.24$& $0.38$& j1-1 &$3.70$&  $2.94$& $2.93$& $4.07$& $4.06$ \\
        c2-2 &$-13.78$&  $-13.29$& $-14.14$& $-12.90$& $-13.76$& f4-2 &$2.48$&  $5.01$& $1.50$& $5.48$& $1.98$& j1-2 &$6.89$&  $6.66$& $6.64$& $7.68$& $7.67$ \\
        c2-3 &$-16.99$&  $-14.32$& $-16.26$& $-14.03$& $-15.98$& f4-3 &$3.33$&  $10.53$& $2.74$& $10.85$& $3.08$& j1-3 &$15.15$&  $15.35$& $15.30$& $16.32$& $16.27$ \\
        c3-1 &$1.09$&  $1.18$& $0.99$& $1.59$& $1.39$& f5-1 &$0.33$&  $0.21$& $-0.47$& $0.53$& $-0.15$& j2-1 &$2.81$&  $3.09$& $2.99$& $3.58$& $3.48$ \\
        c3-2 &$4.36$&  $6.03$& $5.23$& $6.49$& $5.67$& f5-2 &$1.46$&  $3.00$& $0.26$& $3.45$& $0.73$& j2-2 &$7.08$&  $8.06$& $7.62$& $8.37$& $7.94$ \\
        c3-3 &$10.58$&  $14.99$& $13.11$& $15.42$& $13.52$& f5-3 &$1.47$&  $6.60$& $0.68$& $6.86$& $0.97$& j2-3 &$15.35$&  $17.72$& $16.66$& $17.95$& $16.91$ \\
        c4-1 &$5.94$&  $5.57$& $5.27$& $6.08$& $5.77$& g1-1 &$3.60$&  $3.16$& $2.92$& $4.25$& $4.01$& j3-1 &$6.28$&  $10.61$& $7.32$& $10.73$& $7.48$ \\
        c4-2 &$12.83$&  $13.93$& $12.68$& $14.47$& $13.22$& g1-2 &$10.21$&  $11.57$& $10.50$& $12.35$& $11.30$& j3-2 &$25.96$&  $41.96$& $28.75$& $41.44$& $28.37$ \\
        c4-3 &$25.12$&  $30.28$& $27.32$& $30.74$& $27.77$& g1-3 &$21.50$&  $24.99$& $22.63$& $25.65$& $23.31$& j3-3 &$60.88$&  $97.51$& $67.53$& $96.59$& $66.89$ \\
        c5-1 &$0.16$&  $0.39$& $-0.01$& $0.81$& $0.40$& g2-1 &$4.68$&  $6.96$& $5.37$& $7.25$& $5.68$& j4-1 &$1.53$&  $2.64$& $1.21$& $2.96$& $1.55$ \\
        c5-2 &$2.60$&  $5.54$& $3.87$& $5.85$& $4.17$& g2-2 &$16.91$&  $24.72$& $19.02$& $24.67$& $19.00$& j4-2 &$5.99$&  $11.99$& $6.40$& $12.37$& $6.83$ \\
        c5-3 &$12.57$&  $16.63$& $12.85$& $16.92$& $13.12$& g2-3 &$38.67$&  $54.90$& $43.08$& $54.59$& $42.81$& j4-3 &$14.83$&  $28.70$& $16.35$& $28.54$& $16.31$ \\
        d1-1 &$4.73$&  $5.05$& $4.26$& $6.16$& $5.39$& g3-1 &$-0.18$&  $0.40$& $0.05$& $0.72$& $0.37$& j5-1 &$0.83$&  $1.48$& $0.60$& $1.81$& $0.94$ \\
        d1-2 &$11.02$&  $14.42$& $11.55$& $15.47$& $12.65$& g3-2 &$1.18$&  $3.49$& $2.12$& $3.69$& $2.31$& j5-2 &$2.82$&  $6.65$& $3.19$& $6.88$& $3.45$ \\
        d1-3 &$22.93$&  $29.58$& $23.76$& $30.40$& $24.69$& g3-3 &$5.20$&  $10.11$& $7.17$& $10.22$& $7.27$& j5-3 &$7.11$&  $15.93$& $8.39$& $15.99$& $8.52$ \\ 
        d2-1 &$0.12$&  $0.71$& $-0.03$& $1.27$& $0.54$& g4-1 &$0.50$&  $0.66$& $-0.13$& $1.10$& $0.32$& & & & & & \\
        d2-2 &$0.45$&  $3.73$& $0.81$& $4.27$& $1.38$& g4-2 &$1.95$&  $5.10$& $1.90$& $5.42$& $2.23$& & & & & & \\
        d2-3 &$3.72$&  $12.00$& $5.46$& $12.48$& $5.99$& g4-3 &$6.32$&  $13.97$& $6.73$& $14.10$& $6.89$& & & & & & \\\cmidrule(lr){13-18}
        \multicolumn{1}{c}{\vdots} & \multicolumn{1}{c}{\vdots} & \multicolumn{1}{c}{\vdots} & \multicolumn{1}{c}{\vdots} & \multicolumn{1}{c}{\vdots} & \multicolumn{1}{c}{\vdots} & \multicolumn{1}{c}{\vdots} & \multicolumn{1}{c}{\vdots} & \multicolumn{1}{c}{\vdots} & \multicolumn{1}{c}{\vdots} & \multicolumn{1}{c}{\vdots} & \multicolumn{1}{c}{\vdots} & \multicolumn{2}{c}{$E_\mathrm{D} / 10^{-5}$} &$3.28$ & $0.88$ & $3.42$& $0.94$ \\ 
        \bottomrule
   
    \end{tabular}
    
    \label{tab:dragvaluesglobal}
\end{table}

\bibliographystyle{plainnat}
\bibliography{library.bib}

\begin{thebibliography}{27}
\providecommand{\natexlab}[1]{#1}
\providecommand{\url}[1]{\texttt{#1}}
\expandafter\ifx\csname urlstyle\endcsname\relax
  \providecommand{\doi}[1]{doi: #1}\else
  \providecommand{\doi}{doi: \begingroup \urlstyle{rm}\Url}\fi

\bibitem[Abadi et~al.(2016)Abadi, Barham, Chen, Chen, Davis, Dean, Devin,
  Ghemawat, Irving, Isard, Kudlur, Levenberg, Monga, Moore, Murray, Steiner,
  Tucker, Vasudevan, Warden, Wicke, Yu, and Zheng]{TF:2016}
M.~Abadi, P.~Barham, J.~Chen, Z.~Chen, A.~Davis, J.~Dean, M.~Devin,
  S.~Ghemawat, G.~Irving, M.~Isard, M.~Kudlur, J.~Levenberg, R.~Monga,
  S.~Moore, D.~G. Murray, B.~Steiner, P.~Tucker, V.~Vasudevan, P.~Warden,
  M.~Wicke, Y.~Yu, and X.~Zheng.
\newblock {TensorFlow}: A system for {Large-Scale} machine learning.
\newblock In \emph{12th USENIX Symposium on Operating Systems Design and
  Implementation (OSDI 16)}, pages 265--283, Savannah, GA, November 2016.
  USENIX Association.
\newblock ISBN 978-1-931971-33-1.
\newblock URL
  \url{https://www.usenix.org/conference/osdi16/technical-sessions/presentation/abadi}.

\bibitem[Agostini(2020)]{agostini2020exploration}
L.~Agostini.
\newblock Exploration and prediction of fluid dynamical systems using
  auto-encoder technology.
\newblock \emph{Physics of Fluids}, 32\penalty0 (6), 2020.
\newblock ISSN 1070-6631.
\newblock \doi{10.1063/5.0012906}.

\bibitem[Chollet et~al.(2015)]{chollet2015keras}
F.~Chollet et~al.
\newblock Keras.
\newblock \url{https://keras.io}, 2015.

\bibitem[Eivazi et~al.(2020)Eivazi, Veisi, Naderi, and Esfahanian]{eivazi:2020}
H.~Eivazi, H.~Veisi, M.~H. Naderi, and V.~Esfahanian.
\newblock Deep neural networks for nonlinear model order reduction of unsteady
  flows.
\newblock \emph{Physics of Fluids}, 32\penalty0 (10):\penalty0 105104, 2020.
\newblock \doi{10.1063/5.0020526}.

\bibitem[Hou et~al.(2022)Hou, Li, Chen, Wei, Wang, and Huang]{hou2022novel}
Y.~Hou, H.~Li, H.~Chen, W.~Wei, J.~Wang, and Y.~Huang.
\newblock {A novel deep U-Net-LSTM framework for time-sequenced hydrodynamics
  prediction of the SUBOFF AFF-8}.
\newblock \emph{Engineering Applications of Computational Fluid Mechanics},
  16\penalty0 (1):\penalty0 630--645, 2022.
\newblock \doi{10.1080/19942060.2022.2030802}.
\newblock URL \url{https://doi.org/10.1080/19942060.2022.2030802}.

\bibitem[Kingma and Ba(2015)]{Kingma:2015}
D.~P. Kingma and J.~Ba.
\newblock Adam: {A} method for stochastic optimization.
\newblock In \emph{Proceedings of the 3rd International Conference on Learning
  Representations (ICLR)}, 2015.

\bibitem[K{\"u}hl et~al.(2022)K{\"u}hl, Nguyen, Palm, J{\"u}rgens, and
  Rung]{voith22}
N.~K{\"u}hl, T.T. Nguyen, M.~Palm, D.~J{\"u}rgens, and T.~Rung.
\newblock Adjoint node-based shape optimization of free-floating vessels.
\newblock \emph{Structural and Multidisciplinary Optimization}, 2022.
\newblock \doi{10.1007/s00158-022-03338-2}.

\bibitem[Lazzara et~al.(2022)Lazzara, Chevalier, Colombo, {Garay Garcia},
  Lapeyre, and Teste]{lazzara:2022}
M.~Lazzara, M.~Chevalier, M.~Colombo, J.~{Garay Garcia}, C.~Lapeyre, and
  O.~Teste.
\newblock Surrogate modelling for an aircraft dynamic landing loads simulation
  using an lstm autoencoder-based dimensionality reduction approach.
\newblock \emph{Aerospace Science and Technology}, 126:\penalty0 107629, 2022.
\newblock ISSN 1270-9638.
\newblock \doi{https://doi.org/10.1016/j.ast.2022.107629}.
\newblock URL
  \url{https://www.sciencedirect.com/science/article/pii/S1270963822003030}.

\bibitem[Liu et~al.(2011)Liu, Papanikolaou, and
  Zaraphonitis]{liu2011prediction}
S.~Liu, A.~Papanikolaou, and G.~Zaraphonitis.
\newblock Prediction of added resistance of ships in waves.
\newblock \emph{Ocean Engineering}, 38\penalty0 (4):\penalty0 641--650, 2011.

\bibitem[Loft et~al.(2023)Loft, K\"uhl, Buckley, Carpenter, Hinze, Veron, and
  Rung]{loft2023twophase}
M.~Loft, N.~K\"uhl, M.~P. Buckley, J.~R. Carpenter, M.~Hinze, F.~Veron, and
  T.~Rung.
\newblock Two-phase flow simulations of surface waves in wind-forced
  conditions.
\newblock \emph{Physics of Fluids}, 35:\penalty0 072108, 2023.
\newblock \doi{10.1063/5.0156963}.

\bibitem[Lumley(1967)]{lumley1967structure}
J.~L. Lumley.
\newblock The structure of inhomogeneous turbulent flows.
\newblock \emph{Atmospheric turbulence and radio wave propagation}, pages
  166--178, 1967.

\bibitem[Luo-Theilen and Rung(2017)]{luo2017computation}
X.~Luo-Theilen and T.~Rung.
\newblock Computation of {M}echanically {C}oupled {B}odies in a {S}eaway.
\newblock \emph{Ship Technology Research}, 64\penalty0 (3):\penalty0 129--143,
  2017.
\newblock \doi{10.1080/09377255.2017.1348654}.

\bibitem[Luo-Theilen and Rung(2019)]{luo2019numerical}
X.~Luo-Theilen and T.~Rung.
\newblock Numerical {A}nalysis of the {I}nstallation {P}rocedures of {O}ffshore
  {S}tructures.
\newblock \emph{Ocean Engineering}, 179:\penalty0 116--127, 2019.
\newblock \doi{10.1016/j.oceaneng.2019.03.004}.

\bibitem[Maas et~al.(2013)Maas, Hannun, and Ng]{maas:2013}
A.~L. Maas, A.~Y. Hannun, and A.~Y. Ng.
\newblock Rectifier nonlinearities improve neural network acoustic models.
\newblock In \emph{Proceedings of the 30th International Conference on Machine
  Learning}, volume~28, 2013.

\bibitem[Menter(1994)]{menter1994two}
F.R. Menter.
\newblock Two-{E}quation {E}ddy-{V}iscosity {T}urbulence {M}odels for
  {E}ngineering {A}pplications.
\newblock \emph{AIAA Journal}, 32\penalty0 (8):\penalty0 1598--1605, 1994.
\newblock \doi{10.2514/3.12149}.

\bibitem[Mi and Cheng(2025)]{Mi31122025}
B.~Mi and W.~Cheng.
\newblock Intelligent aerodynamic modelling method for steady/unsteady flow
  fields of airfoils driven by flow field images based on modified u-net neural
  network.
\newblock \emph{Engineering Applications of Computational Fluid Mechanics},
  19\penalty0 (1):\penalty0 2440075, 2025.
\newblock \doi{10.1080/19942060.2024.2440075}.
\newblock URL \url{https://doi.org/10.1080/19942060.2024.2440075}.

\bibitem[Milano and Koumoutsakos(2002)]{milano2002neural}
M.~Milano and P.~Koumoutsakos.
\newblock Neural network modeling for near wall turbulent flow.
\newblock \emph{Journal of Computational Physics}, 182\penalty0 (1):\penalty0
  1--26, 2002.

\bibitem[Muzaferija et~al.(1999)Muzaferija, Peric, Sames, and
  Schellin]{muzaferija1999two}
S.~Muzaferija, M.~Peric, P.~Sames, and T.~Schellin.
\newblock A {T}wo-{F}luid {N}avier-{S}tokes {S}olver to {S}imulate {W}ater
  {E}ntry.
\newblock In \emph{Syposium on Naval Hydrodynamics}, pages 638--651. National
  Academy Press, 1999.

\bibitem[Pache and Rung(2022)]{pache2022data}
R.~Pache and T.~Rung.
\newblock Data-driven surrogate modeling of aerodynamic forces on the
  superstructure of container vessels.
\newblock \emph{Engineering applications of computational fluid mechanics},
  16\penalty0 (1):\penalty0 746--763, 2022.

\bibitem[Rung et~al.(2009)Rung, W{\"o}ckner, Manzke, Brunswig, Ulrich, and
  St{\"u}ck]{rung2009challenges}
T.~Rung, K.~W{\"o}ckner, M.~Manzke, J.~Brunswig, C.~Ulrich, and A.~St{\"u}ck.
\newblock Challenges and {P}erspectives for {M}aritime {CFD} {A}pplications.
\newblock \emph{Jahrbuch der Schiffbautechnischen Gesellschaft}, 103:\penalty0
  127--39, 2009.

\bibitem[Schwarz et~al.(2025{\natexlab{a}})Schwarz, Lin, Zemke, and
  Rung]{schwarz2025disentangled}
H.~Schwarz, P.~P. Lin, J.-P.~M. Zemke, and T.~Rung.
\newblock Disentangled latent spaces for reduced order models using
  deterministic autoencoders.
\newblock \emph{arXiv preprint arXiv:2502.14679}, 2025{\natexlab{a}}.
\newblock URL \url{https://arxiv.org/abs/2502.14679}.

\bibitem[Schwarz et~al.(2025{\natexlab{b}})Schwarz, \"{U}berr\"{u}ck, Zemke,
  and Rung]{schwarz:2024}
H.~Schwarz, M.~\"{U}berr\"{u}ck, J.-P.~M. Zemke, and T.~Rung.
\newblock Machine learning based prediction of ditching loads.
\newblock \emph{AIAA Journal}, 63\penalty0 (5):\penalty0 1835--1854,
  2025{\natexlab{b}}.
\newblock \doi{10.2514/1.J064086}.
\newblock URL \url{https://doi.org/10.2514/1.J064086}.

\bibitem[SIMMAN(2008)]{KCS}
SIMMAN.
\newblock {MOERI Container Ship (KCS) - Workshop on Verification and Validation
  of Ship Manoeuvering Simulation Methods}.
\newblock \url{http://www.simman2008.dk/KCS/kcs_geometry.htm}, 2008.
\newblock Accessed: 2023-09-14.

\bibitem[Solera-Rico et~al.(2024)Solera-Rico, Vila, Gómez-López, Wang,
  Almashjary, Dawson, , and Vinuesa]{solera-rico:2024}
A.~Solera-Rico, C.~Sanmiguel Vila, M.~Gómez-López, Y.~Wang, A.~Almashjary,
  S.~Dawson, , and R.~Vinuesa.
\newblock $\beta$-variational autoencoders and transformers for reduced-order
  modelling of fluid flows.
\newblock \emph{Nature Communications}, 15, 02 2024.
\newblock \doi{10.1038/s41467-024-45578-4}.

\bibitem[Swischuk et~al.(2019)Swischuk, Mainini, Peherstorfer, and
  Willcox]{Swischuk.2019}
R.~Swischuk, L.~Mainini, B.~Peherstorfer, and K.~Willcox.
\newblock Projection-based model reduction: Formulations for physics-based
  machine learning.
\newblock \emph{Computers {\&} Fluids}, 179:\penalty0 704--717, 2019.
\newblock ISSN 00457930.
\newblock \doi{https://doi.org/10.1016/j.compfluid.2018.07.021}.

\bibitem[V{\"o}lkner et~al.(2017)V{\"o}lkner, Brunswig, and
  Rung]{volkner2017analysis}
S.~V{\"o}lkner, J.~Brunswig, and T.~Rung.
\newblock Analysis of {N}on-{C}onservative {I}nterpolation {T}echniques in
  {O}verset {G}rid {F}inite-{V}olume {M}ethods.
\newblock \emph{Computers \& Fluids}, 148:\penalty0 39--55, 2017.
\newblock \doi{10.1016/j.compfluid.2017.02.010}.

\bibitem[Wu et~al.(2021)Wu, Gong, Pan, Qiu, Feng, and Pain]{wu:2021}
P.~Wu, S.~Gong, K.~Pan, F.~Qiu, W.~Feng, and C.~Pain.
\newblock Reduced order model using convolutional auto-encoder with
  self-attention.
\newblock \emph{Physics of Fluids}, 33\penalty0 (7):\penalty0 077107, 2021.
\newblock \doi{10.1063/5.0051155}.

\end{thebibliography}

\end{document}